\documentclass[12pt,doublespacing]{article}
\usepackage{fullpage}
\usepackage{setspace}
\usepackage{authblk}
\usepackage{cite}
\usepackage[title]{appendix}
\usepackage{amssymb}
\usepackage{graphicx}
\usepackage{graphics}
\usepackage[cmex10]{amsmath}
\usepackage{latexsym,amsfonts}
\usepackage{amsthm}
\usepackage{subfigure}
\usepackage{hyperref} 
\usepackage{enumitem}

\begin{document}
\title{Coupled Solution of Volume Integral and Hydrodynamic Equations to Analyze Electromagnetic Scattering from Composite Nanostructures}
%\title{Solution of Volume Integral and Hydrodynamic Equations to Analyze Electromagnetic Scattering from Composite Nanostructures}
\author[1]{Doolos~Aibek Uulu}
\author[2]{Rui Chen} 
\author[1]{Liang Chen} 
\author[3]{Ping Li} 
\author[1]{Hakan Bagci} 

\affil[1]{Doolos Aibek Uulu, Liang Chen, and Hakan Bagci are with the Electrical and Computer Engineering Program, Computer, Electrical, and Mathematical Science and Engineering Division, King Abdullah University of Science and Technology (KAUST), Thuwal 23955-6900, Saudi Arabia
\par(e-mail: doolos.aibekuulu@kaust.edu.sa; liang.chen@kaust.edu.sa; hakan.bagci@kaust.edu.sa).\vspace{0.25cm}}

\affil[2]{Rui Chen was with the Electrical and Computer Engineering Program, Computer, Electrical, and Mathematical Science and Engineering Division, King Abdullah University of Science and Technology (KAUST), Thuwal 23955-6900, Saudi Arabia. He is now with the Department of Communication Engineering, Nanjing University of Science and Technology, Nanjing 210094, China
\par (e-mail: rui.chen@kaust.edu.sa).\vspace{0.25cm}}

\affil[3]{Ping Li is with Key Laboratory of Ministry of Education of Design and Electromagnetic Compatibility of High-Speed Electronic Systems, Shanghai Jiao Tong University, Shanghai 200240, China.}

\footnotetext[1]{This work was supported by the King Abdullah University of Science and Technology (KAUST) Office of Sponsored Research (OSR) under Award No 2019-CRG8-4056.}

\date{}

\maketitle
% As a general rule, do not put math, special symbols or citations
% in the abstract or keywords.
\newpage
\begin{abstract}
A coupled system of volume integral and hydrodynamic equations is solved to analyze electromagnetic scattering from nanostructures consisting of metallic and dielectric parts. In the metallic part, the hydrodynamic equation relates the free electron polarization current to the electric flux and effectively ``updates'' the constitutive relation to enable the modeling of nonlocality. In the metallic and the dielectric parts, the volume integral equation relates the electric flux and the free electron polarization current to the scattered electric field. Unknown electric flux and free electron polarization current are expanded using Schaubert-Wilton-Glisson basis functions. Inserting these expansions into the coupled system of the volume integral and hydrodynamic equations and using Galerkin testing yield a matrix system in unknown expansion coefficients. An efficient two-level iterative solver is proposed to solve this matrix system. This approach ``inverts'' the discretized hydrodynamic equation for the coefficients of the free electron polarization current and substitutes the result in the discretized volume integral equation. Outer iterations solve this reduced matrix system while the inner iterations invert the discretized hydrodynamic equation at every iteration of the outer iterations. Numerical experiments are carried out to demonstrate the accuracy, the efficiency, and the applicability of the proposed method.\par
\end{abstract}

\newpage

% For peer review papers, you can put extra information on the cover
% page as needed:
% \ifCLASSOPTIONpeerreview
% \begin{center} \mathbfseries EDICS Category: 3-BBND \end{center}
% \fi
%
% For peerreview papers, this IEEEtran command inserts a page break and
% creates the second title. It will be ignored for other modes.
%\IEEEpeerreviewmaketitle

\section{Introduction}
% The very first letter is a 2 line initial drop letter followed
% by the rest of the first word in caps.
% 
% form to use if the first word consists of a single letter:
% \IEEEPARstart{A}{demo} file is ....
% 
% form to use if you need the single drop letter followed by
% normal text (unknown if ever used by the IEEE):
% \IEEEPARstart{A}{}demo file is ....
% 
% Some journals put the first two words in caps:
% \IEEEPARstart{T}{his demo} file is ....
% 
% Here we have the typical use of a "T" for an initial drop letter
% and "HIS" in caps to complete the first word.
In recent years, with the dramatic advances in fabrication technologies, the use of plasmonic nanostructures to manipulate high-frequency electromagnetic fields has become more prevalent than ever~\cite{ref1,ref2}. Often, metals are used as the building blocks of these nanostructures since they support surface plasmon modes at optical frequencies. These modes localize the electromagnetic fields in the proximity of the nanostructure and significantly enhance those scattered from it in the far-field region. This has enabled the use of metallic nanostructures as nanoantennas~\cite{ref3}, resonators~\cite{ref4}, waveguides~\cite{ref5}, couplers~\cite{ref6}, and sensors~\cite{ref7}. 

Depending on the frequency, interaction of electromagnetic fields with metals can be accounted for using various models and equations under certain assumptions and approximations. At microwave frequencies, free electrons in a metal have high mobility, which leads to large conductivity and small skin depth (compared to the size of the structure)~\cite{balanis2015antenna}. Therefore, an electric current, which is confined to the surface of the metal, is used to represent the electromagnetic field interactions on the metal. At optical frequencies, the free electron mobility decreases. As a result, there is a time delay in the response of the electrons to the electromagnetic excitation~\cite{maier2007plasmonics}. To account for this frequency dispersion effect, classical Drude model~\cite{dressel2006verifying} is used to represent the permittivity of the metal. Furthermore, at the optical frequencies, the skin depth is usually comparable to the size of a typical nanostructure and the metals are modeled as ``penetrable’’ materials and volume electric currents are used to represent the electromagnetic field interactions on or inside them. 

When the frequency is further increased into the ultraviolet part of the spectrum, spatial dispersion appears in the response of the free electrons to the electromagnetic excitation. Effectively, the permittivity becomes nonlocal, i.e., it depends not only on the observation point but also the source point in space~\cite{forstmann2006metal}. This spatial dispersion effect is due to the fact that, at this frequency regime, a free electron exhibits quantum behavior~\cite{mortensen2013nonlocal}, and the interactions of the electromagnetic fields with the electrons should ideally be modeled using full quantum mechanics simulation methods (e.g., density functional theory~\cite{marques2006time}). However, the computational cost of these methods is very high, and, therefore, they can only be used when the structure is very small, i.e., only a few nanometers in size~\cite{teperik2013robust}. This makes them unsuitable for full-scale simulations of plasmonic nanostructures in real-life scenarios.

This bottleneck can be addressed by using a hydrodynamic equation to model the mechanical motion of the free electrons~\cite{forstmann2006metal}. This equation assumes that the electrons can collectively be accounted for as a moving fluid of charges driven by the electromagnetic fields in the medium. Naturally, these moving charges (which is termed as the free electron polarization current in the rest of the text) generate electromagnetic fields. This interaction between the free electrons and the electromagnetic fields can be described by a coupled system of the Maxwell and the hydrodynamic equations~\cite{zheng2018boundary, hiremath2012numerical, bhardwaj2018fast, schmitt2018simulation, li2017hybridizable}. This system of coupled equations can account for the nonlocality/spatial dispersion and its numerical solution is not as costly as that of the full quantum mechanics simulation methods.

Several methods have been developed to numerically solve the coupled system of the Maxwell and the hydrodynamic equations~\cite{zheng2018boundary, hiremath2012numerical, bhardwaj2018fast, schmitt2018simulation, li2017hybridizable}. Majority of these methods are differential equation solvers, e.g., finite element method~\cite{hiremath2012numerical}, finite-difference time-domain method~\cite{bhardwaj2018fast}, and discontinuous Galerkin method~\cite{schmitt2018simulation, li2017hybridizable}. These solvers, just like their traditional versions, which are developed to solve only the Maxwell equations,  suffer from several well-known shortcomings that might limit their accuracy and efficiency (see for example~\cite{jin2011theory} for details). %For open region scattering problems, they require the unbounded physical domain to be truncated into a bounded computation domain that includes the metallic scatterer. This is often done using absorbing boundary conditions or perfectly matched layers that introduce errors in the solution. This error can be controlled but this comes with increased computational requirements. Differential equation solvers discretize the whole computation domain using volumetric elements resulting in a large number of unknowns to be solved for. 

Surface integral equation solvers do not suffer from these shortcomings (see for example again~\cite{jin2011theory} for details), and indeed, they have been extended to
%They implicitly enforce the radiation condition and do not require absorbing boundary conditions or perfectly matched layers since a Green function is used to represent the scattered electromagnetic fields. Integral equation solvers discretize only the surface or the volume of the scatterer resulting in a smaller number of unknowns (compared to differential equation solvers). Indeed, to take advantage of these benefits, in~\cite{zheng2018boundary, zheng2019review}, a surface integral equation solver has been developed 
analyze scattering from metallic objects in which the motion of electrons is described by the hydrodynamic equation\cite{zheng2018boundary}. However, this method requires the derivation of a new Green function for every type of boundary condition and their combination enforced by the hydrodynamic equation (for example boundary conditions for the free electron polarization current on metal-metal or metal-dielectric interfaces are different). In addition, this solver is applicable only when the scatterer has homogeneous or piece-wise homogeneous material properties. 

In this work, these shortcomings are avoided by switching to a volume integral equation formulation. The proposed scheme represents the scattered electromagnetic field in the form of a (volumetric) convolution between the background medium’s Green function and the electric flux (induced in the metallic and dielectric parts of the scatterer) and the free electron polarization current (induced in the metallic part). In both the metallic and the dielectric parts, the volume integral equation, which relates the electric flux and the free electron polarization current to the scattered electric field, is enforced. In the metallic part, the hydrodynamic equation, which relates the free electron polarization current to the electric flux, is enforced. To numerically solve this coupled system of equations, first the scatterer is discretized into a mesh of tetrahedrons. The electric flux and the free electron polarization current are expanded using a combination of ``full'' and ``half'' Schaubert-Wilton-Glisson (SWG) basis functions~\cite{schaubert1984tetrahedral} defined on these tetrahedrons. Inserting these expansions into the coupled system of the volume integral and the hydrodynamic equations and applying Galerkin testing yield a matrix system in unknown expansion coefficients. The boundary condition for the normal component of the free electron polarization current (which should be set to zero on a metal-dielectric interface) is enforced by excluding the half SWG functions, which are defined on the tetrahedrons that have a face on the metal-dielectric interface, from the expansion of the free electron polarization current. 

An efficient two-level iterative solver is developed to
solve the matrix system resulting from this discretization. This solver ``inverts'' the discretized hydrodynamic equation for the coefficients of the free electron polarization current and substitutes the result in the discretized volume integral equation. Outer iterations solve this reduced
matrix system while the inner iterations invert the discretized
hydrodynamic equation at every iteration of the outer iterations. Note that a preliminary version of the method proposed in this work has been described in~\cite{uulu2020} as a conference contribution. 

The remainder of this paper is organized as follows. Sections~\ref{sec:form},~\ref{sec:disc}, and~\ref{sec:solu} describe the formulation of the coupled system of the volume integral and the hydrodynamic equations, its discretization, and the two-level iterative solution, respectively. Section~\ref{sec:comm} explains how the mesh element size is selected and provides several comments on possible extensions of the proposed method and its applications. Section~\ref{sec:numr} presents numerical results that demonstrate the accuracy, the efficiency, and the applicability of the proposed scheme. Finally, Section~\ref{sec:conc} summarizes this work and outlines future research directions. 

\section{Formulation}
\subsection{Coupled System of the Volume Integral and the Hydrodynamic Equations}\label{sec:form}
Let $V_\mathrm{D}$ represent a composite scatterer that consists of dielectric and metallic parts (Fig.~\ref{fig:scat}). $V_{\mathrm{diel}}$ and $V_{\mathrm{H}}$ represent these parts, respectively. Both $V_{\mathrm{diel}}$ and $V_{\mathrm{H}}$ are non-magnetic but their relative permittivity (for bound electron polarization) as denoted by $\varepsilon_{\mathrm{b}}(\mathbf{r})$ can be inhomogeneous. The boundary surface of $V_{\mathrm{H}}$ is represented by $S_{\mathrm{H}}$. The scatterer resides in an unbounded homogeneous background medium with permittivity ${\varepsilon _0}$ and permeability $\mu _0$. The electric field of the excitation is denoted by ${{\mathbf{E}}^{{\mathrm{inc}}}}({\mathbf{r}})$ and its frequency is denoted by $\omega$. Upon this excitation, equivalent volumetric electric current ${\mathbf{J}}({\mathbf{r}})$ is induced in $V_\mathrm{D}$ and this current generates the scattered electric field ${{\mathbf{E}}^{{\mathrm{sca}}}}({\mathbf{r}})$. The incident electric field ${{\mathbf{E}}^{{\mathrm{inc}}}}({\mathbf{r}})$, the scattered electric field $\mathbf{E}^{\mathrm{sca}}(\mathbf{r})$, and the total electric field $\mathbf{E}(\mathbf{r})$ satisfy the fundamental field relation:
\begin{figure}[!t]
\centering
\includegraphics[width=0.6\columnwidth]{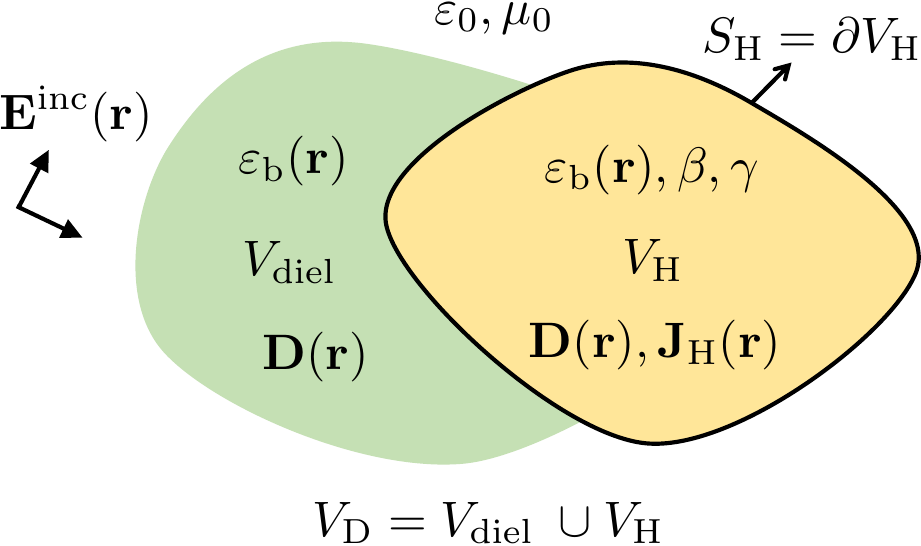}
\caption{Description of the scattering problem.}
\label{fig:scat}
\end{figure}
\begin{equation}
\label{eq9}
    {\mathbf{E}}({\mathbf{r}}) = {{\mathbf{E}}^{{\mathrm{inc}}}}({\mathbf{r}}) + {{\mathbf{E}}^{{\mathrm{sca}}}}({\mathbf{r}}).
\end{equation}
Using the volume equivalence principle~\cite{jin2011theory}, ${\mathbf{J}}({\mathbf{r}})$ is expressed in terms of ${\mathbf{E}}({\mathbf{r}})$ and the electric flux ${\mathbf{D}}({\mathbf{r}})$ as
\begin{equation}
\label{eq7}
       {\mathbf{J}}({\mathbf{r}}) = j\omega {\mathbf{D}}({\mathbf{r}}) - j\omega {\varepsilon _0}{\mathbf{E}}({\mathbf{r}}),\;\mathbf{r}\in V_\mathrm{D}
\end{equation}
and $\mathbf{E}^{\mathrm{sca}}(\mathbf{r})$ is expressed as
\begin{equation}
\label{eq6}
    {{\mathbf{E}}^{{\mathrm{sca}}}}({\mathbf{r}})=\mathcal{L}_{V_{\mathrm{D}}}[\mathbf{J}](\mathbf{r}).
\end{equation}
Here, $\mathcal{L}_{V}[\mathbf{X}](\mathbf{r})$ is the volume integral operator given by
\begin{equation*}
\begin{aligned}
    \mathcal{L}_{V}[\mathbf{X}](\mathbf{r})=&-j \omega \mu_{0} \int_{V} \mathbf{X}\left(\mathbf{r}^{\prime}\right) G\left(\mathbf{r}, \mathbf{r}^{\prime}\right) dv^{\prime} \\
&+\frac{1}{j \omega \varepsilon_{0}}  \int_{V}\nabla \nabla \cdot [\mathbf{X}\left(\mathbf{r}^{\prime}\right) G\left(\mathbf{r}, \mathbf{r}^{\prime}\right)] dv^{\prime}
\end{aligned}
\end{equation*}
where $G({\mathbf{r}},{\mathbf{r'}}) = \exp ( - j{k_0}R)/(4\pi R)$ is the Green function and ${k_0} = \omega \sqrt {{\mu _0}{\varepsilon _0}}$ is the wavenumber in the background medium, and $R = \left| {{\mathbf{r}} - {\mathbf{r'}}} \right|$ denotes the distance between observation point ${\mathbf{r}}$ and source point ${\mathbf{r'}}$. 

The motion of the free electrons in $V_\mathrm{H}$ is accounted for using a nonlocal hydrodynamic equation. Let $\mathbf{J}_{\mathrm{H}}(\mathbf{r})$ represent the polarization current associated with these free electrons. Then, $\mathbf{J}_{\mathrm{H}}(\mathbf{r})$ is expressed in terms of $\mathbf{E}(\mathbf{r})$ and $\mathbf{D}(\mathbf{r})$ as~\cite{zheng2018boundary}
\begin{equation}
\label{eq4}
      {{{\mathbf{J}}_{{\mathrm{H}}}}({\mathbf{r}})} = {j\omega }{\mathbf{D}}({\mathbf{r}})-{j\omega }{\varepsilon _0}{\varepsilon _{\mathrm{b}}(\mathbf{r})}{\mathbf{E}}({\mathbf{r}}),\;\mathbf{r}\in V_\mathrm{H}.
\end{equation}
In addition, $\mathbf{J}_{\mathrm{H}}(\mathbf{r})$ is ``driven’’ by $\mathbf{E}(\mathbf{r})$. This relationship is described by the nonlocal hydrodynamic Drude equation as~\cite{forstmann2006metal}
\begin{equation}
\label{eq3}
\begin{aligned}
{\beta ^2}\nabla \left[ {\nabla  \cdot {{\mathbf{J}}_{{\mathrm{H}}}}({\mathbf{r}})} \right] &+ \omega \left( {\omega  - j\gamma } \right){{\mathbf{J}}_{{\mathrm{H}}}}({\mathbf{r}}) \\ &= - j\omega \omega^2_{\mathrm{p}}{\varepsilon _0}{\mathbf{E}}({\mathbf{r}}),\;\mathbf{r}\in V_\mathrm{H}. 
\end{aligned}
\end{equation}
Here, ${\omega_\mathrm{p}}$ is the plasma frequency, ${\beta^2} = 0.6v_{\mathrm{F}}^2$ , $v_\mathrm{F}$ is the Fermi velocity, and $\gamma $ is a damping constant. The hydrodynamic equation~\eqref{eq3} has to be complemented by a boundary condition on the boundary surface of $V_{\mathrm{H}}$ [as denoted by $S_{\mathrm{H}}$ (Fig.~\ref{fig:scat})]~\cite{forstmann2006metal}:
\begin{equation}
\label{eq5}
    {\mathbf{\hat n}}(\mathbf{r}) \cdot {{\mathbf{J}}_{{\mathrm{H}}}}({\mathbf{r}}) = 0,\;\mathbf{r}\in S_\mathrm{H}.
\end{equation}
Here, ${\mathbf{\hat n}}(\mathbf{r})$ is the outward pointing unit normal vector on $S_{\mathrm{H}}$. This boundary condition ensures that the normal component of $ \mathbf{J}_{\mathrm{H}}(\mathbf{r})$ vanishes on $S_{\mathrm{H}}$ (i.e., the free electrons do not flow from the metallic part into the dielectric part or the background medium). Note that the hydrodynamic equation~\eqref{eq3} together with the boundary condition~\eqref{eq5} introduces an electromagnetic wave solution with a longitudinal electric field in $V_{\mathrm{H}}$ in addition to the one with a transverse electric field~\cite{forstmann2006metal}.

The proposed scheme defines ${\mathbf{D}}({\mathbf{r}})$ and ${{\mathbf{J}}_{{\mathrm{H}}}}({\mathbf{r}})$ as unknowns. To eliminate ${\mathbf{E}}({\mathbf{r}})$, ${\mathbf{E}}({\mathbf{r}})$ from~\eqref{eq4} is inserted into~\eqref{eq7}. This yields an expression for $\mathbf{J}(\mathbf{r})$ in terms of ${\mathbf{D}}({\mathbf{r}})$ and ${{\mathbf{J}}_{{\mathrm{H}}}}({\mathbf{r}})$ as
\begin{equation}
\label{eq8}
        {\mathbf{J}}({\mathbf{r}}) = j\omega \kappa(\mathbf{r}){\mathbf{D}}({\mathbf{r}}) + \frac{{{\mathbf{J}}_{{\mathrm{H}}}}({\mathbf{r}})}{{\varepsilon _{\mathrm{b}}(\mathbf{r})}},\;\mathbf{r}\in V_\mathrm{H}
\end{equation}
where $\kappa(\mathbf{r}) = 1-1/{\varepsilon _{\mathrm{b}}(\mathbf{r})}$. Substituting~\eqref{eq8} into~\eqref{eq6} and inserting the resulting equation and ${\mathbf{E}}({\mathbf{r}})$ from~\eqref{eq4} into~\eqref{eq9} yield the volume integral equation in unknowns ${\mathbf{D}}({\mathbf{r}})$ and ${{\mathbf{J}}_{{\mathrm{H}}}}({\mathbf{r}})$ as
\begin{equation}
\label{eq10}
\begin{aligned}
{{\mathbf{E}}^{{\mathrm{inc}}}}({\mathbf{r}}) &= \frac{{{\mathbf{D}}({\mathbf{r}})}}{{{\varepsilon _0}{\varepsilon _{\mathrm{b}}(\mathbf{r})}}} - j\omega\mathcal{L}_{V_{\mathrm{D}}}[\mathbf{\kappa\mathbf{D}}](\mathbf{r})\\
&-\frac{\mathbf{J}_{\mathrm{H}}(\mathbf{r})}{j \omega \varepsilon_{0} \varepsilon_{\mathrm{b}}(\mathbf{r})}-\mathcal{L}_{V_{\mathrm{H}}}[\frac{\mathbf{J}_{\mathrm{H}}}{\varepsilon_{\mathrm{b}}}](\mathbf{r}),\; \mathbf{r} \in V_{\mathrm{D}}.
%\nonumber{{\mathbf{E}}^{{\mathrm{inc}}}}({\mathbf{r}}) &= \frac{{{\mathbf{D}}({\mathbf{r}})}}{{{\varepsilon _0}{\varepsilon _{\mathrm{b}}(\mathbf{r})}}} - \frac{{{{\mathbf{J}}_{{\mathrm{H}}}}({\mathbf{r}})}}{{j\omega {\varepsilon _0}{\varepsilon _{\mathrm{b}}(\mathbf{r})}}}\\&
% \nonumber-\omega^2 {\mu _0}\int_{V_\mathrm{D}} { \kappa(\mathbf{r'}){\mathbf{D}}({\mathbf{r'}})} G({\mathbf{r}},{\mathbf{r'}})d\mathbf{r}'\\&
% \nonumber+ j\omega {\mu _0}\int_{V_\mathrm{H}} {{\mathbf{J}}_{{\mathrm{H}}}}({\mathbf{r'}})/{\varepsilon _{\mathrm{b}}(\mathbf{r'})} G({\mathbf{r}},{\mathbf{r'}})d\mathbf{r}'\\&
% \nonumber - \frac{1}{{{\varepsilon _0}}}\nabla \int_{V_\mathrm{D}} {\nabla ' \cdot\{ \kappa(\mathbf{r'}){\mathbf{D}}({\mathbf{r'}})} \}G({\mathbf{r}},{\mathbf{r'}})d\mathbf{r}' \\&- \frac{1}{{j\omega {\varepsilon _0}}}\nabla \int_{V_\mathrm{H}} \nabla ' \cdot \{ {{\mathbf{J}}_{{\mathrm{H}}}}({\mathbf{r'}})/{\varepsilon _{\mathrm{b}}(\mathbf{r'})}\} G({\mathbf{r}},{\mathbf{r'}})d\mathbf{r}',\;\;\;\;\mathbf{r}\in V_\mathrm{H}. 
 \end{aligned}
\end{equation}
Similarly, the hydrodynamic equation~\eqref{eq3} should be expressed in only ${\mathbf{D}}({\mathbf{r}})$ and ${{\mathbf{J}}_{{\mathrm{H}}}}({\mathbf{r}})$. Inserting ${\mathbf{E}}({\mathbf{r}})$ from~\eqref{eq4} into~\eqref{eq3} yields 
\begin{equation}
\label{eq11}
\begin{aligned}
 {\beta ^2}\nabla \left[ {\nabla  \cdot {{\mathbf{J}}_{{\mathrm{H}}}}({\mathbf{r}})} \right]& + \left[ {\omega (\omega  - j\gamma ) - \frac{{\omega _\mathrm{p}^2}}{{{\varepsilon _{\mathrm{b}}(\mathbf{r})}}}} \right]{{\mathbf{J}}_{{\mathrm{H}}}}({\mathbf{r}}) \\&=  - j\omega \omega^2_{\mathrm{p}}\frac{{{\mathbf{D}}({\mathbf{r}})}}{{{\varepsilon _{\mathrm{b}}(\mathbf{r})}}},\;\mathbf{r}\in V_\mathrm{H}.
\end{aligned}
\end{equation}
Eqs.~\eqref{eq10} and~\eqref{eq11} are the final form of the coupled system of the volume integral and the hydrodynamic equations in unknowns ${\mathbf{D}}({\mathbf{r}})$ and ${{\mathbf{J}}_{{\mathrm{H}}}}({\mathbf{r}})$. This system is discretized using the scheme described in Section~\ref{sec:disc}. Note that this discretization scheme ensures that the boundary condition~\eqref{eq5} is enforced correctly. 

\subsection{Discretization}\label{sec:disc}
To numerically solve the coupled system of~\eqref{eq10} and~\eqref{eq11}, first, $V_\mathrm{D}$ is discretized into a mesh of tetrahedrons. Then, the unknowns ${\mathbf{D}}({\mathbf{r}})$ and ${{\mathbf{J}}_{{\mathrm{H}}}}({\mathbf{r}})$ are expanded as
\begin{equation}
\label{eq12}
\begin{aligned}
&{\mathbf{D}}({\mathbf{r}}) = \sum\limits_{n = 1}^{{N_{\mathrm{D}}}} {{{\{ {{\mathbf{I}}_{{\mathrm{D}}}}\} }_n}{\mathbf{f}}_n^{\mathrm{D}}({\mathbf{r}})},\;\mathbf{r}\in V_\mathrm{D} \\
&{{\mathbf{J}}_{{\mathrm{H}}}}({\mathbf{r}}) = \mathop \sum \limits_{n = 1}^{{N_{{\mathrm{H}}}}} {\{ {{\mathbf{I}}_{{\mathrm{H}}}}\} _n}{\mathbf{f}}_n^{{\mathrm{H}}}({\mathbf{r}}),\;\mathbf{r}\in V_\mathrm{H}.
\end{aligned}
\end{equation}
Here, ${\{ {{\mathbf{I}}_{{\mathrm{D}}}}\} _n}$ and ${\{ {{\mathbf{I}}_{{\mathrm{H}}}}\} _n}$ are the unknown coeffcients and ${\mathbf{f}}_n^{\mathrm{D}}({\mathbf{r}})$ and ${\mathbf{f}}_n^{{\mathrm{H}}}({\mathbf{r}})$ are the basis functions constructed using the SWG functions defined on the triangles of the tetrahedral mesh~\cite{schaubert1984tetrahedral}. The SWG basis function associated with triangle $n$ is defined as
\begin{equation}
\label{eq:swg}
\mathbf{f}_{n}(\mathbf{r})=\left\{\begin{aligned}
&\mathbf{f}_{n}^{+}(\mathbf{r}) = \frac{|S_{n}|}{3|V_{n}^{+}|}(\mathbf{r}-\mathbf{r}_{n}^{+}),\; \mathbf{r}\in V_{n}^{+} \\
&\mathbf{f}_{n}^{-}(\mathbf{r}) = -\frac{|S_{n}|}{3|V_{n}^{-}|}(\mathbf{r}-\mathbf{r}_{n}^{-}),\; \mathbf{r}\in V_{n}^{-} \\
& 0, \mathrm{elsewhere}
\end{aligned}\right..
\end{equation}
Here, $V_n^+$ and $V_n^-$ are the tetrahedrons ``touching'' triangle $n$ on its two sides, ${{\bf r}}_n^ \pm $ are the corners of $V_n^ \pm $ that are not on $S_n$ (i.e., free nodes), $\left| {{S_n}} \right|$ is the area of ${S_n}$, and $\left| {V_n^ \pm } \right|$ are the volumes of $V_n^ \pm $. 

The basis set ${\mathbf{f}}_n^{\mathrm{D}}({\mathbf{r}})$ includes ``full’’ SWG basis functions as defined by~\eqref{eq:swg} on every pair of tetrahedrons in $V_{\mathrm{D}}$ as well as ``half'' SWG basis functions defined by $\mathbf{f}_{n}^{+}(\mathbf{r})$ of~\eqref{eq:swg} in single tetrahedrons that have their $S_n$ on the surface of the scatterer. The use of full SWG functions enforces the continuity of the normal component of ${\mathbf{D}}({\mathbf{r}})$ across any pair of tetrahedrons in $V_{\mathrm{D}}$ (even when $\varepsilon_{\mathrm{b}}(\mathbf{r})$ in $V_n^+$ and $V_n^-$ are different). The inclusion of half SWG functions ensures that the normal component of ${\mathbf{D}}({\mathbf{r}})$ on the surface of the scatterer is accounted for. ${N_{\mathrm{D}}}$ in~\eqref{eq12} is the total number of full and half SWG basis functions used in the expansion of ${\mathbf{D}}({\mathbf{r}})$ in $V_{\mathrm{D}}$.

The basis set ${\mathbf{f}}_n^{\mathrm{H}}({\mathbf{r}})$ consists of only the full basis functions as defined by~\eqref{eq:swg} on every pair of tetrahedrons in $V_{\mathrm{H}}$ and does \emph{not} include the half SWG basis functions defined in single tetrahedrons that have their $S_n$ on $S_{\mathrm{H}}$ (metal-dielectric interface). Note that ${{\mathbf{J}}_{{\mathrm{H}}}}({\mathbf{r}})$ does not flow from the metallic part into the dielectric part or the background medium and its normal component on $S_{\mathrm{H}}$ is zero as described by the boundary condition in~\eqref{eq5}. Exclusion of the half SWG basis functions from ${\mathbf{f}}_n^{{\mathrm{H}}}({\mathbf{r}})$ ensures that this boundary condition is correctly enforced. 
${N_{{\mathrm{H}}}}$ in~\eqref{eq12} is the total number of full SWG basis functions used in the expansion of ${\mathbf{J}}_{\mathrm{H}}({\mathbf{r}})$ in $V_{\mathrm{H}}$.

Inserting the expansions~\eqref{eq12} into~\eqref{eq10} and~\eqref{eq11} and Galerkin testing the resulting equations using ${\mathbf{f}}_m^{\mathrm{D}}({\mathbf{r}})$, $m = 1,2,...,{N_{\mathrm{D}}}$  and ${\mathbf{f}}_m^{{\mathrm{H}}}({\mathbf{r}})$, $m = 1,2,...,{N_{{\mathrm{H}}}}$ yield a coupled matrix system of dimension $({N_{\mathrm{D}}} + {N_{{\mathrm{H}}}}) \times ({N_{\mathrm{D}}} + {N_{{\mathrm{H}}}})$ as
\begin{equation}
\label{eq14}
\underbrace{\begin{bmatrix}
{{{\mathbf{Z}}_{{\mathrm{DD}}}}}&{{{\mathbf{Z}}_{{{\mathrm{DH}}}}}}\\
{{{\mathbf{Z}}_{{{\mathrm{HD}}}}}}&{{{\mathbf{Z}}_{{{\mathrm{HH}}}}}}
\end{bmatrix}}_{\displaystyle \mathbf{Z}}
\underbrace{\begin{bmatrix}
{{{\mathbf{I}}_{{\mathrm{D}}}}}\\
{{{\mathbf{I}}_{{\mathrm{H}}}}}
\end{bmatrix}}_{\displaystyle \mathbf{I}} = \underbrace{\begin{bmatrix}
{{{\mathbf{V}}^{{\mathrm{inc}}}}}\\
{\mathbf{0}}
\end{bmatrix}}_{\displaystyle \mathbf{V}}. 
\end{equation}
In~\eqref{eq14}, the entries of the block matrices $\mathbf{Z}_{\mathrm{DD}}$ of dimension $N_{\mathrm{D}}\times N_{\mathrm{D}}$, $\mathbf{Z}_{\mathrm{DH}}$ of dimension $N_{\mathrm{D}}\times N_{\mathrm{H}}$, $\mathbf{Z}_{\mathrm{HD}}$ of dimension $N_{\mathrm{H}}\times N_{\mathrm{D}}$, $\mathbf{Z}_{\mathrm{HH}}$ of dimension $N_{\mathrm{H}}\times N_{\mathrm{H}}$, and the tested incident field vector $\mathbf{V}^\mathrm{inc}$ of dimension $N_\mathrm{D}$ are given by
\begin{equation}
\label{eq:ZDD}
\begin{gathered}
\{\mathbf{Z}_{\mathrm{DD}}\}_{mn}=\frac{1}{\varepsilon_0}\Big\langle\mathbf{f}_{m}^{\mathrm{D}}(\mathbf{r}),\frac{\mathbf{f}_n^{\mathrm{D}}(\mathbf{r})}{\varepsilon_{\mathrm{b}}(\mathbf{r})}\Big\rangle\\
-j \omega \Big\langle\mathbf{f}_{m}^{\mathrm{D}}(\mathbf{r}),\mathcal{L}_{V_{\mathrm{D}}}[\kappa \mathbf{f}_{n}^{\mathrm{D}}](\mathbf{r})\Big\rangle\\
\end{gathered}
\end{equation}
\begin{equation}
\label{eq:ZDH}
\begin{gathered}
\{\mathbf{Z}_{\mathrm{DH}}\}_{mn}=-\frac{1}{j\omega\varepsilon_0}\Big\langle\mathbf{f}_{m}^{\mathrm{D}}(\mathbf{r}),\frac{\mathbf{f}_n^{\mathrm{H}}(\mathbf{r})}{\varepsilon_{\mathrm{b}}(\mathbf{r})}\Big\rangle\\
-\Big\langle\mathbf{f}_{m}^{\mathrm{D}}(\mathbf{r}),\mathcal{L}_{V_{\mathrm{H}}}[\frac{\mathbf{f}_{n}^{\mathrm{H}}}{\varepsilon_{\mathrm{b}}}](\mathbf{r})\Big\rangle
\end{gathered}
\end{equation}
\begin{equation}
\label{eq:ZHD}
\{\mathbf{Z}_{\mathrm{HD}}\}_{mn}=j \omega \omega_{\mathrm{p}}^{2}\Big\langle\mathbf{f}_{m}^{\mathrm{H}}(\mathbf{r}),\frac{\mathbf{f}_{n}^{\mathrm{D}}(\mathbf{r})}{\varepsilon_{\mathrm{b}}(\mathbf{r})}\Big\rangle
\end{equation}
\begin{equation}
\label{eq:ZHH}
\begin{gathered}
\{\mathbf{Z}_{\mathrm{HH}}\}_{mn}=\beta^{2}\Big\langle\mathbf{f}_{m}^{\mathrm{H}}(\mathbf{r}),\nabla\left[\nabla \cdot \mathbf{f}_{n}^{\mathrm{H}}(\mathbf{r})\right]\Big\rangle\\
+\omega(\omega-j \gamma)\Big\langle \mathbf{f}_{m}^{\mathrm{H}}(\mathbf{r}),\mathbf{f}_{n}^{\mathrm{H}}(\mathbf{r})\Big\rangle-\omega_{\mathrm{p}}^{2}\Big\langle\mathbf{f}_{m}^{\mathrm{H}}(\mathbf{r}),\frac{\mathbf{f}_{n}^{\mathrm{H}}(\mathbf{r})}{\varepsilon_{\mathrm{b}}(\mathbf{r})}\Big\rangle
\end{gathered}
\end{equation}
\begin{equation}
\label{eq:Vinc}
\{\mathbf{V}^{\mathrm{inc}}\}_{m}=\Big\langle\mathbf{f}_{m}^{\mathrm{D}}(\mathbf{r}),\mathbf{E}^{\mathrm{inc}}(\mathbf{r})\Big\rangle
\end{equation}
respectively. Here, the inner product between vector functions $\mathbf{a}(\mathbf{r})$ and $\mathbf{b}(\mathbf{r})$ is defined as
\begin{equation}
    \Big\langle \mathbf{a}(\mathbf{r}),\mathbf{b}(\mathbf{r})\Big\rangle = \int_{V_{a}}\mathbf{a}(\mathbf{r})\cdot\mathbf{b}(\mathbf{r})\,dv
\end{equation}
where $V_{a}$ is the support of $\mathbf{a}(\mathbf{r})$. Since $\mathcal{L}_{V_{\mathrm{D}}}(\mathbf{X})[\mathbf{r}]$ and $\mathcal{L}_{V_{\mathrm{H}}}(\mathbf{X})[\mathbf{r}]$ are ``global'' operators, one can see from~\eqref{eq:ZDD} and~\eqref{eq:ZDH} that $\mathbf{Z}_{\mathrm{DD}}$ and $\mathbf{Z}_{\mathrm{DH}}$ are dense matrix blocks. On the other hand, since $\mathbf{f}_{n}^{\mathrm{D}}(\mathbf{r})$ and $\mathbf{f}_{n}^{\mathrm{H}}(\mathbf{r})$ have ``local'' supports (two tetrahedrons for a full SWG function and one tetrahedron for a half SWG function), one can see from~\eqref{eq:ZHD} and~\eqref{eq:ZHH} that $\mathbf{Z}_{\mathrm{HD}}$ and $\mathbf{Z}_{\mathrm{HH}}$ are sparse matrix blocks (the maximum number of non-zero entries in one row of these blocks is seven). Note that the detailed expressions for the matrix entries in~\eqref{eq:ZDD}-\eqref{eq:ZHH} and the vector entries in~\eqref{eq:Vinc} are provided in the Appendix.

\subsection{Solution of the Matrix Equation}\label{sec:solu}
To take advantage of the sparsity of $\mathbf{Z}_{\mathrm{HD}}$ and $\mathbf{Z}_{\mathrm{HH}}$ directly, the coupled matrix system~\eqref{eq14} is solved iteratively for the unknown coefficient vectors ${{\mathbf{I}}_{\mathrm{D}}}$ and ${{\mathbf{I}}_{\mathrm{H}}}$. Two approaches can be used for this purpose. 

\subsubsection{Single-level iterative solver}
The coupled system~\eqref{eq14} is iteratively solved as a whole using a transpose-free quasi-minimal residual (TFQMR) scheme~\cite{freund1993transpose}. The computational cost of matrix-vector multiplication $\mathbf{Z}\tilde{\mathbf{I}}$ required at every iteration of TFQMR scales as $ \mathcal{O}(N_{\mathrm{D}}^2)+\mathcal{O}(N_{\mathrm{D}}N_{\mathrm{H}})+\mathcal{O}(N_{\mathrm{H}})+\mathcal{O}(N_{\mathrm{H}}) $. The four terms in this expression represent the computational cost of multiplying matrix blocks $\mathbf{Z}_{\mathrm{DD}}$, $\mathbf{Z}_{\mathrm{DH}}$, $\mathbf{Z}_{\mathrm{HD}}$, and $\mathbf{Z}_{\mathrm{HH}}$ with the relevant part of $\tilde{\mathbf{I}}$, respectively. Then, the overall computational cost of this single-level iterative solver scales as 
\begin{equation}
\label{eq20}
\begin{gathered}
\mathcal{O}(N_{\mathrm{it}}N_{\mathrm{D}}^2)+\mathcal{O}(N_{\mathrm{it}}N_{\mathrm{D}}N_{\mathrm{H}})+\mathcal{O}(N_{\mathrm{it}}N_{\mathrm{H}})\\
+\mathcal{O}(N_{\mathrm{it}}N_{\mathrm{H}})
\end{gathered}
\end{equation}
where $N_{\mathrm{it}}$ is the number of iterations required for the relative residual error to converge to a user defined value. 

\subsubsection{Two-level iterative solver} 
In this approach, before an iterative solver is used, the coupled system~\eqref{eq14} is first reduced into a smaller matrix system. This is done by inverting the second row of~\eqref{eq14} for ${{\mathbf{I}}_{\mathrm{H}}}$, i.e., ${{\mathbf{I}}_{\mathrm{H}}} =  - {\mathbf{Z}}_{\mathrm{HH}}^{ - 1}{{\mathbf{Z}}_{\mathrm{HD}}}{{\mathbf{I}}_{\mathrm{D}}}$, and inserting this expression into the first row. This yields a smaller matrix system of dimension ${N_{\mathrm{D}}} \times {N_{\mathrm{D}}}$ in unknown ${{\mathbf{I}}_{\mathrm{D}}}$ as
\begin{equation}
\label{eq22}
({{\mathbf{Z}}_{\mathrm{DD}}} - {{\mathbf{Z}}_{\mathrm{DH}}}{\mathbf{Z}}_{\mathrm{HH}}^{ - 1}{{\mathbf{Z}}_{\mathrm{HD}}}){{\mathbf{I}_{\mathrm{D}}}} = {{\mathbf{V}}^{{\mathrm{inc}}}}.
\end{equation}
Then, TFQMR is used to solve~\eqref{eq22} for ${{\mathbf{I}}_{\mathrm{D}}}$. Matrix-vector multiplication $({{\mathbf{Z}}_{\mathrm{DD}}} - {{\mathbf{Z}}_{\mathrm{DH}}}{\mathbf{Z}}_{\mathrm{HH}}^{ - 1}{{\mathbf{Z}}_{\mathrm{HD}}}){\mathbf{\tilde I_{\mathrm{D}}}}$ required at every iteration of TFQMR is carried out as described below.

\vspace{10pt}
\hrule
\vspace{1pt}
\hrule
\vspace{5pt}
\emph{Step 1}: Compute the first term ${{\mathbf{Z}}_{\mathrm{DD}}}{\mathbf{\tilde I_{\mathrm{D}}}}$.

\emph{Step 2}: Compute the second term ${{\mathbf{Z}}_{\mathrm{DH}}}{\mathbf{Z}}_{\mathrm{HH}}^{ - 1}{{\mathbf{Z}}_{\mathrm{HD}}}{\mathbf{\tilde I_{\mathrm{D}}}}$ in three steps as

\hspace{10pt} \emph{Step 2.1}: Compute $\mathbf{y}=\mathbf{Z}_{\mathrm{HD}} {\mathbf{\tilde I_{\mathrm{D}}}}$. 

\hspace{10pt} \emph{Step 2.2}: Compute $\mathbf{x}=\mathbf{Z}_{\mathrm{HH}}^{-1}\mathbf{y}$ by solving $\mathbf{y}=\mathbf{Z}_{\mathrm{HH}}\mathbf{x}$ for $\mathbf{x}$. This is done iteratively using TFQMR.

\hspace{10pt} \emph{Step 2.3}: Compute ${{\mathbf{Z}}_{\mathrm{DH}}}{\mathbf{x}}$.

\emph{Step 3:} Substract the result of \emph{Step 2.3} from that of \emph{Step 1}.
\vspace{5pt}
\hrule
\vspace{1pt}
\hrule
\vspace{10pt}

\noindent Computational costs of \emph{Step 1}, \emph{Step 2.1}, \emph{Step 2.2}, and \emph{Step 2.3} scale as $\mathcal{O}(N_{\mathrm{D}}^2)$, $\mathcal{O}(N_{\mathrm{H}})$, $\mathcal{O}(N_{\mathrm{it}}^{\mathrm{in}}N_{\mathrm{H}})$, and $\mathcal{O}(N_{\mathrm{D}}N_{\mathrm{H}})$, respectively. Here, $N_{\mathrm{it}}^{\mathrm{in}}$ is the number of iterations required for the relative residual error to converge to a user defined value during the solution of the matrix equation $\mathbf{y}=\mathbf{Z}_{\mathrm{HH}}\mathbf{x}$ at \emph{Step 2.2} (i.e., inner iterations). Then, the overall computational cost of this two-level iterative solver scales as 
\begin{equation}
\label{eq23}
\begin{gathered}
\mathcal{O}(N_{\mathrm{it}}^{\mathrm{out}}N_{\mathrm{D}}^2)+\mathcal{O}(N_{\mathrm{it}}^{\mathrm{out}}N_{\mathrm{H}})+\mathcal{O}(N_{\mathrm{it}}^{\mathrm{out}}N_{\mathrm{it}}^{\mathrm{in}}N_{\mathrm{H}})\\
+\mathcal{O}(N_{\mathrm{it}}^{\mathrm{out}}N_{\mathrm{D}}N_{\mathrm{H}})
\end{gathered}
\end{equation}
where $ N_{\mathrm{it}}^{\mathrm{out}}$ is the number of iterations required for the relative residual error to converge to a user defined value during the solution of the matrix equation~\eqref{eq22} (i.e., outer iterations). 

Comparing~\eqref{eq23} to~\eqref{eq20}, one can see that the single-level iterative solver would certainly be faster than the two-level iterative solver for $N_{\mathrm{it}} \leq N_{\mathrm{it}}^{\mathrm{out}}$. However, numerical results presented in Section~\ref{sec:sphere} show that $N_{\mathrm{it}}^{\mathrm{out}}$ is much smaller than $N_{\mathrm{it}}$ and $ N_{\mathrm{it}}^{\mathrm{in}}$ is small, and therefore the two-level iterative solver is significantly faster than the single-level iterative solver. The difference between $N_{\mathrm{it}}$ and $N_{\mathrm{it}}^{\mathrm{out}}$ can be explained by the fact that the dimension of~\eqref{eq22} is almost half of that of~\eqref{eq14} and inserting  ${{\mathbf{I}}_{\mathrm{H}}} =  - {\mathbf{Z}}_{\mathrm{HH}}^{ - 1}{{\mathbf{Z}}_{\mathrm{HD}}}{{\mathbf{I}}_{\mathrm{D}}}$ into the first row of~\eqref{eq14} [to obtain~\eqref{eq22}] effectively takes care of the scaling difference between the volume integral and the hydrodynamic equations. 

Note that for both approaches, the computational cost of matrix-vector multiplications $\mathbf{Z}_{\mathrm{DD}} {\mathbf{\tilde I_{\mathrm{D}}}}$ and $\mathbf{Z}_{\mathrm{DH}} {\mathbf{\tilde I_{\mathrm{H}}}}$ can be reduced using the fast multipole method~\cite{coifman1993fast,greengard1998accelerating,engheta1992fast,abduljabbar2019extreme} and its multi-level versions~\cite{song1997multilevel,sheng1998solution,jarvenpaa2013broadband,takrimi2017incomplete} as well as other matrix compression schemes like those described and referred to in~\cite{guo2017butterfly,sayed2021butterfly,kaplan2022fast,brick2020increasing}. But this does not change the conclusions of the above comparison since the difference in the computational cost of the two iterative solvers is mainly due to the difference in the number of iterations. 

\subsection{Comments}\label{sec:comm}

\begin{figure}[t]
\centering
\includegraphics[width=0.6\columnwidth]{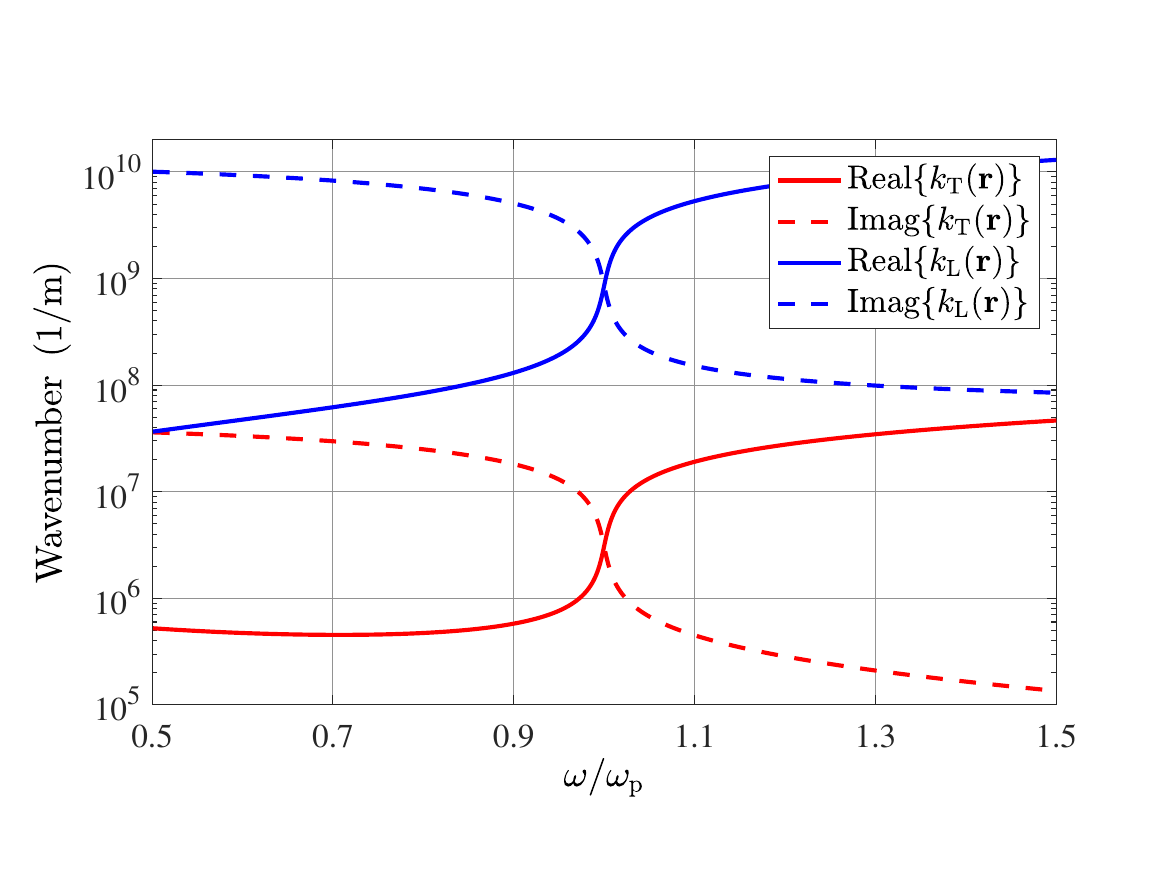}
\caption{The wavenumbers of the electromagnetic waves with the transverse and the longitudinal electric fields, ${k_{\mathrm{T}}}({\mathbf{r}})$ and ${k_{\mathrm{L}}}({\mathbf{r}})$, versus $\omega /{\omega _{\mathrm{p}}}$.}
\label{fig:wavenum}
\end{figure}

A metallic medium that is described by the hydrodynamic equation~\eqref{eq3} supports propagation of electromagnetic waves with electric fields along the transverse and longitudinal directions~\cite{forstmann2006metal}. Let the (complex) wavenumbers associated with these electromagnetic waves be denoted by $k_{\mathrm{T}}(\mathbf{r})$ and $ k_{\mathrm{L}}(\mathbf{r})$, respectively. The expressions of $k_{\mathrm{T}}(\mathbf{r})$ and $ k_{\mathrm{L}}(\mathbf{r})$ are given by~\cite{forstmann2006metal}
\begin{equation}
\label{eq24}
\begin{gathered}
k_{\mathrm{T}}(\mathbf{r})=k_0\sqrt{\varepsilon_{\mathrm{b}}(\mathbf{r})-\frac{\omega_{\mathrm{p}}^2}{\omega^2-j\omega\gamma}}\\
%\label{eq25}
k_{\mathrm{L}}(\mathbf{r})=\frac{1}{\beta}\sqrt{\omega^2-j\omega\gamma-\frac{\omega_{\mathrm{p}}^2}{\varepsilon_{\mathrm{b}}(\mathbf{r})}}.
\end{gathered}
\end{equation}
For example, for gold, $\varepsilon _{\mathrm{b}}(\mathbf{r})= 1$,  $\omega_{\mathrm{p}}=1.20 \times {10^{16}}\,{\mathrm{s^{-1}}}$,  $\beta=1.07 \times{10^{6}}\,\mathrm{m/s}$,  and $\gamma=1.36 \times {10^{14}}\,{\mathrm{s^{-1}}}$~\cite{rakic98,kreiter2002surface}. Fig.~\ref{fig:wavenum} plots the values of $k_{\mathrm{T}}(\mathbf{r})$ and $ k_{\mathrm{L}}(\mathbf{r})$ computed for gold in the frequency range $\omega \in [0.5{\omega _{\mathrm{p}}}, 1.5{\omega _{\mathrm{p}}}]$. Note that during the computation of square roots in~\eqref{eq24}, the positive imaginary part of the result is selected to avoid a non-physical growing wave. The figure clearly shows that in this frequency range, both real and imaginary parts of ${k_{\mathrm{L}}}({\mathbf{r}})$ are significantly larger than those of ${k_{\mathrm{T}}}({\mathbf{r}})$, respectively. This means that to accurately capture the behaviour of the electromagnetic fields in $V_{\mathrm{H}}$ (inside the metallic part), the mesh of tetrahedrons must resolve the wavelength associated with ${k_{\mathrm{L}}}({\mathbf{r}})$. Note that within the frequency range considered here, since $k_0$ is significantly smaller than both ${k_{\mathrm{L}}}({\mathbf{r}})$ and ${k_{\mathrm{T}}}({\mathbf{r}})$, the mesh in $V_{\mathrm{diel}}= V_{\mathrm{D}}- V_{\mathrm{H}}$ (inside the dielectric part) can ideally be coarser than the one in $V_{\mathrm{H}}$. But since these two volumes share a surface and a conformal discretization is used, the mesh in $V_{\mathrm{diel}}$ is denser than what it would ideally be. This unnecessary computational overhead can be alleviated by switching to a non-conformal discretization scheme, such as the one described in~\cite{zhang2017discontinuous,kong2018discontinuous}. Development of such a scheme is underway. 
 
Several comments about the possible extensions of the proposed method are in order:
\begin{enumerate}
\item The formulation in Section~\ref{sec:form} assumes that the scatterer includes only one type of homogeneous metalic structure (described by a set of hydrodynamic equation parameters $\gamma$, $\beta$, and $\omega_{\mathrm{p}}$). The formulation can easily account for additional metal types described by assigning different values to $\gamma$, $\beta$, and $\omega_{\mathrm{p}}$  in relevant matrix entries as long as the structures made of different metals do not touch each other. This limitation stems from the fact that the boundary condition~\eqref{eq5} is not valid on metal-metal interfaces and therefore the discretization scheme described in Section ~\ref{sec:disc} is not applicable anymore. Development of a formulation that removes this limitation and allows for modeling of metal-metal interfaces is currently underway.
\item In~\cite{mortensen2014generalized}, the nonlocal hydrodynamic Drude model [as described mathematically in~\eqref{eq3}] has been extended to account for the classical kinetic effects of the charge carrier diffusion in the nonlocal response. This so-called generalized nonlocal optical response (GNOR) model expresses the relationship between $\mathbf{J}_{\mathrm{H}}(\mathbf{r})$ and $\mathbf{E}(\mathbf{r})$ as 
\begin{align}
\nonumber \left[\beta^2+D(\gamma+j \omega)\right]&\nabla \left[ {\nabla  \cdot {{\mathbf{J}}_{{\mathrm{H}}}}({\mathbf{r}})} \right] + \omega \left( {\omega  - j\gamma } \right){{\mathbf{J}}_{{\mathrm{H}}}}({\mathbf{r}}) \\
\label{eq30}= - j\omega& \omega^2_{\mathrm{p}}{\varepsilon _0}{\mathbf{E}}({\mathbf{r}}),\;\mathbf{r}\in V_\mathrm{H}.
\end{align}
Here, $\beta$, $\gamma$, and $\omega_{\mathrm{p}}$ are same as those in~\eqref{eq3} and the additional parameter $D$ is the charge carrier diffusion constant. Comparing~\eqref{eq3} and~\eqref{eq30}, one can easily see that the only difference between the two equations is the additional term $D(\gamma+j \omega)$ in~\eqref{eq30}. Furthermore, both the nonlocal models use the same boundary condition given in~\eqref{eq5}. Therefore, the implementation of the GNOR model within the numerical scheme proposed here is rather trivial and can simply be done by replacing $\beta^2$ by $\beta^2+D(\gamma+j\omega)$.
\item In~\cite{christensen2014nonlocal,zouros2020monitoring}, an analytical method that relies on the Mie series expansion of the fields has been used to study the nonlocal response of nanospheres in three different scenarios: electromagnetic scattering, electron energy-loss spectroscopy, and atomic spontaneous emission. Even though examples presented in Section~\ref{sec:numr} involve only electromagnetic scattering problems under plane-wave excitation, the proposed numerical scheme is applicable to other problems with different types of excitation, including electron energy-loss and atomic spontaneous emission.
\end{enumerate}

\section{Numerical Results}\label{sec:numr}
In this section, several numerical examples are presented to demonstrate the accuracy, the efficiency, and the applicability of the proposed solver. All scatterers considered in these examples reside in free space with permittivity $\varepsilon_0$ and permeability $\mu_0$. For all the examples, the excitation is a plane wave with electric field 
\begin{equation}
\label{eq26}
    {{\mathbf{E}}^{{\mathrm{inc}}}}({\mathbf{r}}) = {\mathbf{\hat p}}{E_0}{e^{ - j{k_0}\hat{\mathbf{k}}^{\mathrm{inc}} \cdot {\mathbf{r}}}}.
\end{equation}
Here, the unit vector ${\mathbf{\hat p}}$ represents the direction of the polarization, ${E_0}$ is the electric field amplitude, and the unit vector $\hat{\mathbf{k}}^{\mathrm{inc}}$ represents the direction of propagation. Unless otherwise stated, ${\mathbf{\hat p}} = {\mathbf{\hat x}}$, ${E_0} = 1\,{\mathrm{ V/m}}$, and $\hat{\mathbf{k}}^{\mathrm{inc}} = {\mathbf{\hat z}}$ for all examples considered here. All TFQMR iterations (outer and inner iterations for the two-level iterative solver and the iterations for the single-level iterative solver) are terminated when the relative residual error (RRE) reaches a desired level, i.e., when the condition  $|| \mathbf{b}-\mathbf{A}\mathbf{I}_{n}||/||\mathbf{b}|| \leq \chi_{\mathrm{RRE}}$ is satisfied.. Here, ${{\mathbf{I}}_n}$ is the solution at iteration step $n$, $\mathbf{A}$ is the matrix, $\mathbf{b}$ is the right-hand side vector, and ${\chi _{{\mathrm{RRE}}}}$ is the convergence threshold. 
\begin{table}[!t]
\centering
\renewcommand{\arraystretch}{1.3}
\setlength{\tabcolsep}{3pt}
\caption{Performance of the Single-level and the Two-level Iterative Solvers in Analyzing Scattering from a Gold Nanosphere}
\label{tab:perf}
\begin{tabular}{ c | c | c | c | c | c | c }
\multicolumn{2}{c|} {Mesh} & \multicolumn{2}{c|} {Single-level} & \multicolumn{3}{c}{Two-level} \vspace{-0.12cm} \\
\multicolumn{2}{c|} {levels} & \multicolumn{2}{c|} {iterative solver} & \multicolumn{3}{c}{iterative solver} \\ \hline\hline
$N_{\mathrm{D}}$ & $N_{\mathrm{H}}$ & $\mathrm{Time\,(s)}$ & $N_{\mathrm{it}}$ & $\mathrm{Time\,(s)}$ & $N^{\mathrm{out}}_{\mathrm{it}}$ & $N^{\mathrm{in}}_{\mathrm{it}}$ \\ \hline
$5\,494$ & $5\,030$ & $29$ & $84$ & $10$  & $29$ & $30-35$ \\ \hline
$9\,216$ & $8\,612$ & $105$ & $108$ & $32$ & $26$ & $35-40$ \\ \hline
$17\,546$ & $16\,694$ & $530$ & $126$ & $106$ & $24$ & $45-50$\\ \hline \hline
\end{tabular}
\end{table}
\begin{figure}[t!]
\centering
\subfigure[]{\includegraphics[width=0.5\columnwidth]{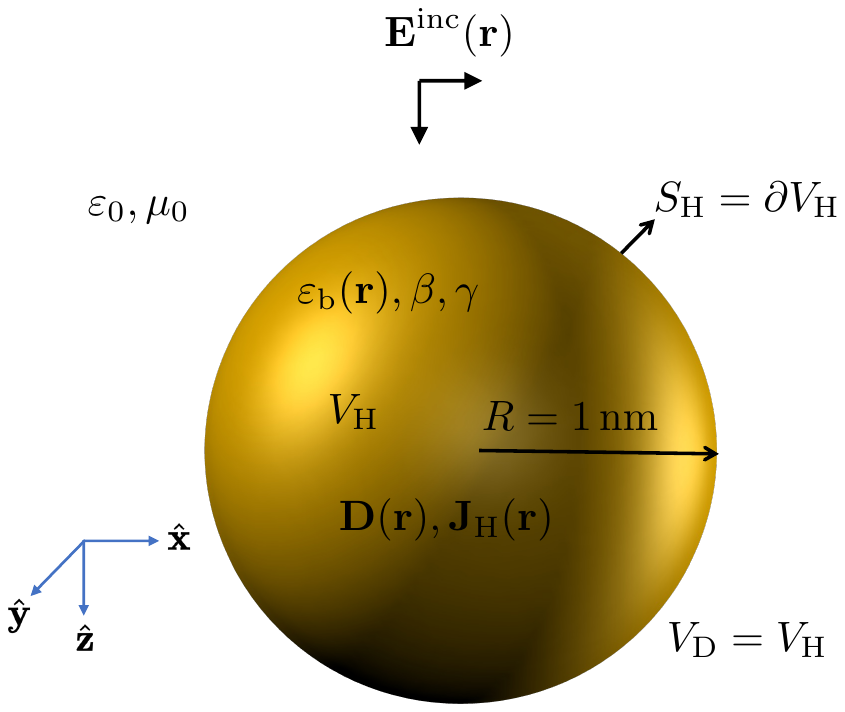}}\\
\subfigure[]{\includegraphics[width=0.495\columnwidth]{{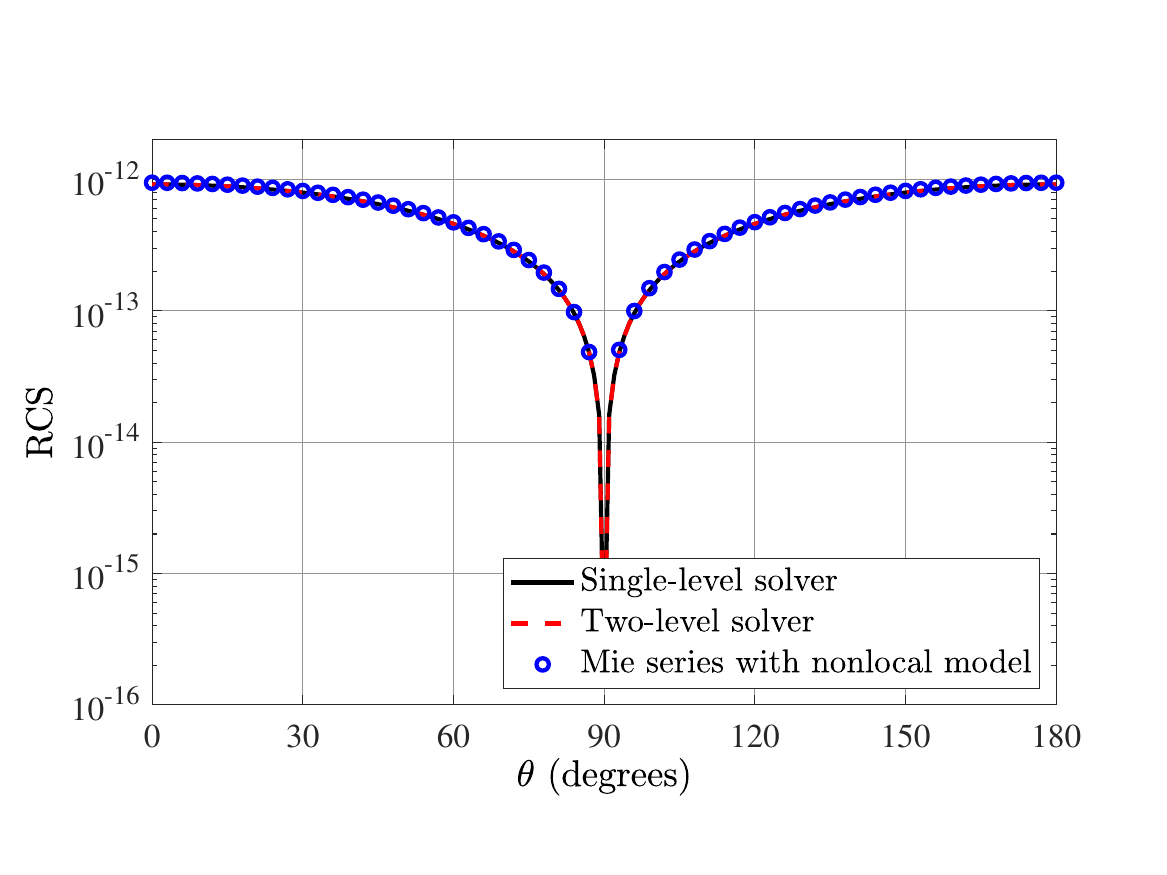}}}
\subfigure[]{\includegraphics[width=0.495\columnwidth]{{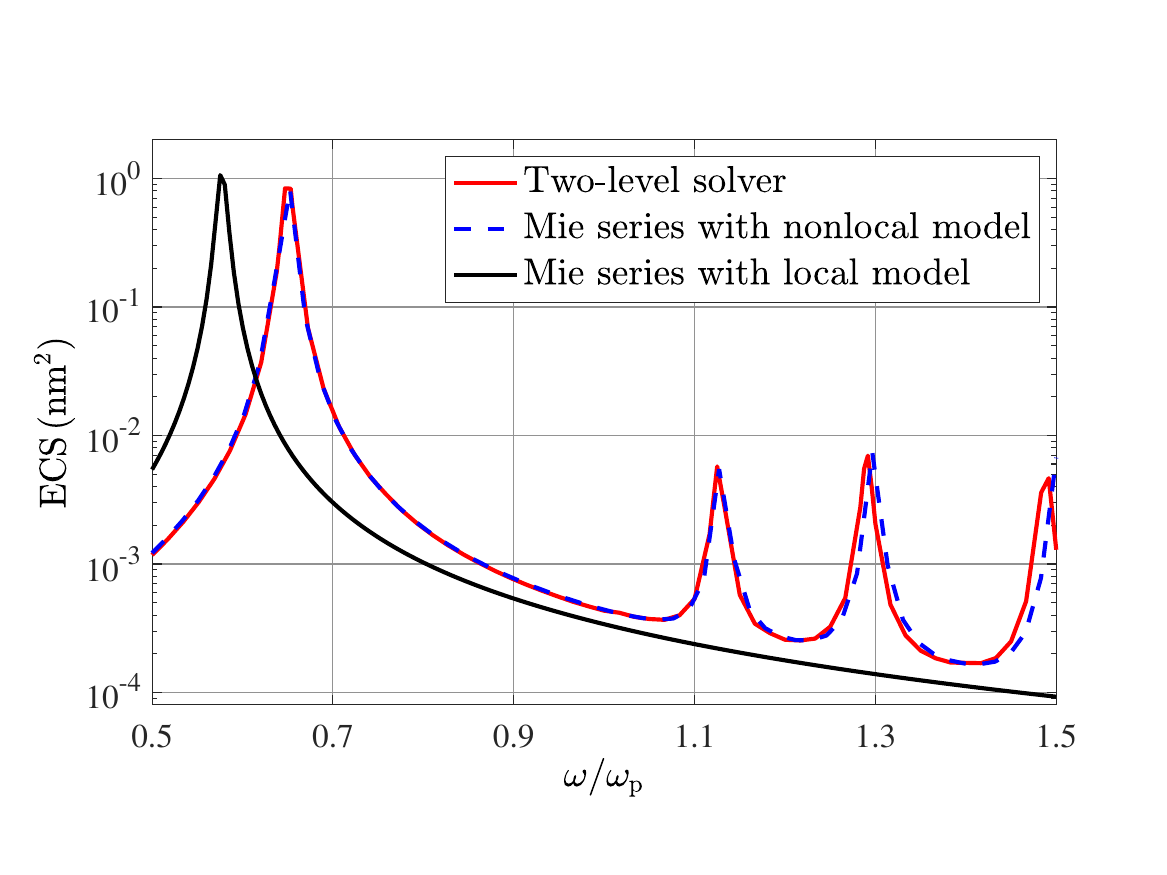}}}
\caption{(a) Description of the scattering problem from involving a gold nanosphere. (b) RCS computed on the $xy$-plane at $\omega = 0.5\,\omega_{\mathrm{p}}$ from the solutions obtained using the single-level and the two-level iterative schemes and the Mie series with nonlocal-response material. (c) ECS computed from the solutions obtained using the two-level iterative scheme and the Mie series with nonlocal- and local-response material models versus $\omega /{\omega_\mathrm{p}}$.}
\label{fig_sphere}
\end{figure}
\subsection{Metallic Sphere}\label{sec:sphere}
In this example, electromagnetic scattering from a gold nanosphere [Fig.~\ref{fig_sphere}(a)] is analyzed using the proposed method. The radius of the sphere is $1\,\mathrm{nm}$. The hydrodynamic equation parameters for gold are ${\omega_{\mathrm{p}}} = 1.20 \times {10^{16}}\,{\mathrm{s^{-1}}}$, $\gamma  = 1.36 \times {10^{14}}\,{\mathrm{s^{-1}}}$, ${v_{\mathrm{F}}} = 1.39 \times {10^6}\,{\mathrm{ m/s}}$, and $\varepsilon_{\mathrm{b}}(\mathbf{r})=1$~\cite{rakic98}. Note that, for this problem, $ V_{\mathrm{D}}=V_{\mathrm{H}}$ and $ V_{\mathrm{diel}} = \emptyset$. 

Two sets of simulations are carried out. For the first set of simulations, frequency is set to $\omega = 0.5\,{\omega _{\mathrm{p}}}$ and three levels of mesh are used. These meshes use $N_{\mathrm{D}}=\{5\,494, 9\,216, 17\,546\}$ and $N_{\mathrm{H}}=\{5\,030, 8\,612, 16\,694\}$ basis functions to discretize $\mathbf{D}(\mathbf{r})$ and $\mathbf{J}_{\mathrm{H}}(\mathbf{r})$ induced inside the sphere, respectively. The single-level and the two-level iterative schemes are used to solve the matrix system~\eqref{eq14}, which is preconditioned from left using a diagonal preconditioner, and the matrix system~\eqref{eq22}, respectively. For the iterations of the single-level scheme and the outer iterations of the two-level scheme, ${\chi_{{\mathrm{RRE}}}}= 10^{-4}$. For the inner iterations of the two-level scheme, ${\chi_{{\mathrm{RRE}}}}= 10^{-8}$. Note that the accuracy of the inner iterations has to be high to ensure that the outer iterations converge. This does not increase the computational cost significantly because the converge rate of the inner iterations is already very fast.

Fig.~\ref{fig_sphere}(b) plots the radar cross section (RCS) computed on the $xy$-plane from the solutions obtained by the single-level and the two-level iterative schemes for the mesh with $N_{\mathrm{D}}= 17\,546$ and $N_{\mathrm{H}}=16\,694$ and the analytical Mie series with nonlocal-response material model~\cite{jackson1999classical,ruppin1973optical}. The results agree very well. Table~\ref{tab:perf} compares the performance of the single-level and the two-level iterative solvers. The two-level iterative solver is significantly faster due to fact that $N_{\mathrm{it}}^{\mathrm{out}}$ is much smaller than $N_{\mathrm{it}}$ and $N_{\mathrm{it}}^{\mathrm{in}}$ is small (see the computational complexity comparison in Section~\ref{sec:solu}). Note that numbers for $N_{\mathrm{it}}^{\mathrm{in}}$ presented in Table I are the range of the inner iterations.

For the second set of simulations, a total of 60 simulations are carried out using the two-level iterative solver at equally spaced points in the frequency range $\omega \in [0.5{\omega _{\mathrm{p}}}, 1.5{\omega _{\mathrm{p}}}]$. In these simulations, $\mathbf{D}(\mathbf{r})$ and $ \mathbf{J}_{\mathrm{H}}(\mathbf{r})$ induced inside the sphere are discretized using $N_{\mathrm{D}}=143\,254$ and ${N_{{\mathrm{H}}}} = 140\,022$ basis functions, respectively. For the outer and the inner iterations of the two-level iterative scheme, ${\chi_{{\mathrm{RRE}}}}= 10^{-4}$ and ${\chi_{{\mathrm{RRE}}}}= 10^{-8}$, respectively.

Fig.~\ref{fig_sphere}(c) plots the extinction cross section (ECS) computed from the solutions obtained using the two-level iterative scheme and the analytical Mie series with nonlocal- and local-response material models versus $\omega /{\omega _{\mathrm{p}}}$~\cite{jackson1999classical,ruppin1973optical}. The figure clearly shows that the result obtained using the proposed method matches well with the result obtained using the Mie series with the nonlocal-response material model. The figure also shows that the error in the result obtained by the proposed method increases with $\omega$. This error can be reduced by using a denser mesh that can capture the behavior of the fields with large $k_{\mathrm{L}}(\mathbf{r})$ more accurately (see Section~\ref{sec:comm} and Fig.~\ref{fig:wavenum}).

The first peak observed in all three ECS curves is caused by the transverse field resonance. Also, a ``blue shift’’ phenomenon is shown in this figure, i.e., the transverse field resonance peak shifts towards higher frequencies when the nonlocal material response is taken into account~\cite{mortensen2014generalized}. Furthermore, three other peaks are identified at $\omega = 1.13\,\omega_{\mathrm{p}}$, $\omega = 1.29\,\omega_{\mathrm{p}}$, and $\omega = 1.50\,\omega_{\mathrm{p}}$  in ECS computed by the proposed solver and the Mie series with the nonlocal-response material model. These are caused by the longitudinal field resonance. It also can be concluded from Fig.~\ref{fig_sphere}(c) that the transverse field response is more dominant at $\omega  < {\omega _{\mathrm{p}}}$ while the longitudinal field response is more dominant at $\omega  > {\omega _{\mathrm{p}}}$.
\subsection{Metallic Dimer}
\begin{figure}[t]
\centering
\subfigure[]{\includegraphics[width=0.6\columnwidth]{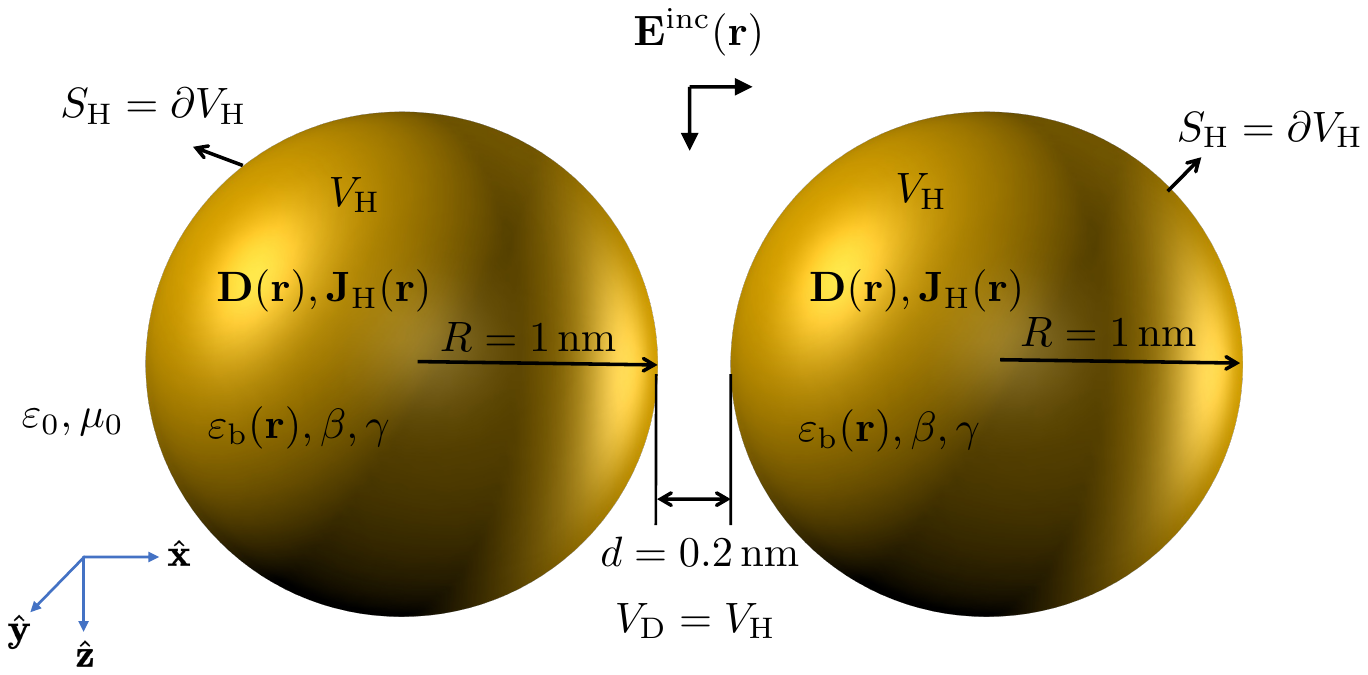}}\\
\subfigure[]{\includegraphics[width=0.6\columnwidth]{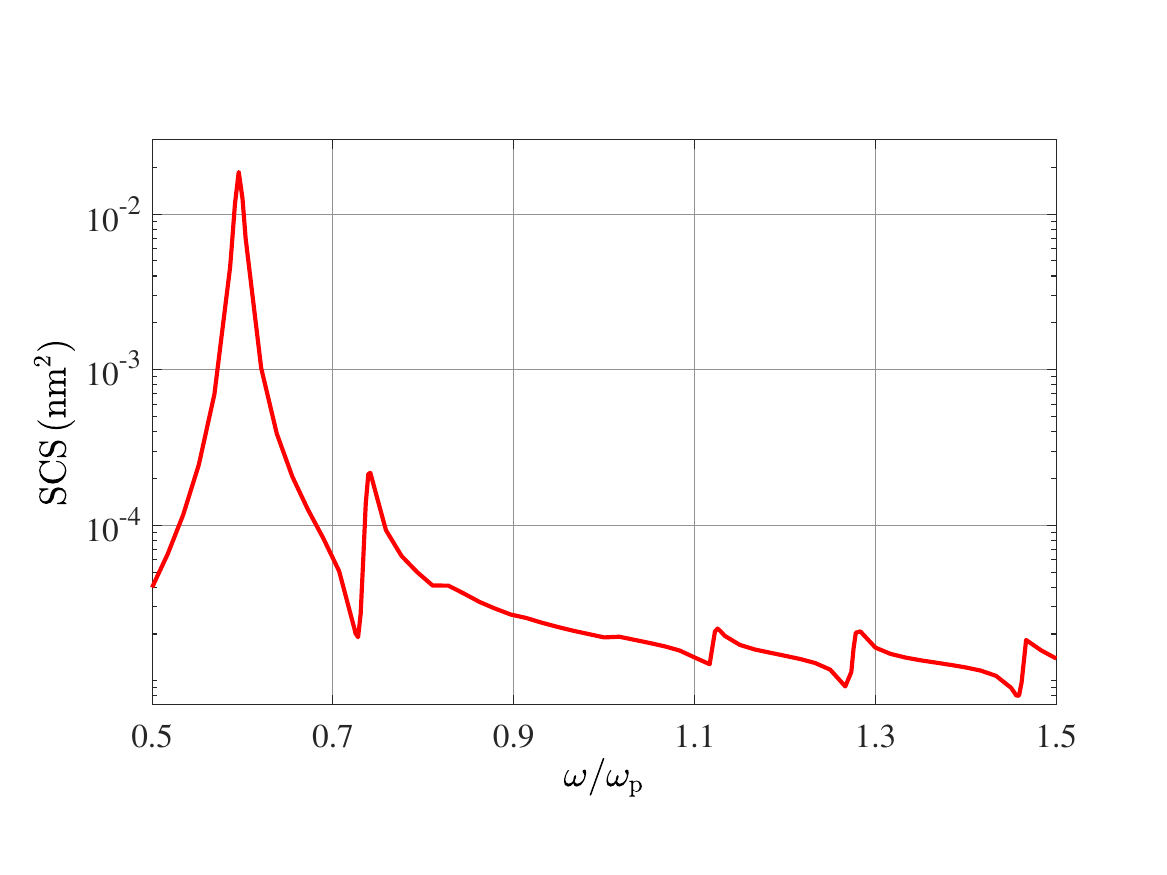}}\\
\caption{(a) Description of the scattering problem involving a nanodimer. (b) SCS computed from the solutions obtained using the two-level iterative scheme versus $\omega /{\omega _\mathrm{p}}$.}
\label{fig_dimer}
\end{figure}

In this example, electromagnetic scattering from a nanodimer [Fig.~\ref{fig_dimer}(a)] is analyzed using the proposed solver. The radius of the spheres is $1\,\mathrm{nm}$ and the shortest distance between them is $0.2\,\mathrm{nm}$. The hydrodynamic equation parameters for the material making up the spheres are ${\omega _{\mathrm{p}}} = 1.20 \times {10^{16}}\,{\mathrm{s^{-1}}}$, $\gamma  = 1.36 \times {10^{14}}\,{\mathrm{s^{-1}}}$, ${v_{\mathrm{F}}} = 1.39 \times {10^6}\,{\mathrm{m/s}}$, and $\varepsilon_{\mathrm{b}}(\mathbf{r})=1$. Note that, for this problem, $ V_{\mathrm{D}}=V_{\mathrm{H}}$ and $ V_{\mathrm{diel}} = \emptyset$. 

A total of $60$ simulations are carried out using the two-level iterative scheme at equally spaced points in the frequency range $\omega \in [0.5 \omega_{\mathrm{p}}, 1.5 \omega_{\mathrm{p}}]$. Both spheres use the same mesh of tetrahedrons and two levels of mesh are used. For simulations at $\omega  < {\omega _{\mathrm{p}}}$, ${N_{\mathrm{D}}} = 104\,194$ and ${N_{{\mathrm{H}}}} = 100\,706$ while for simulations at $\omega  > {\omega _{\mathrm{p}}}$, ${N_{\mathrm{D}}} = 286\,958$ and ${N_{{\mathrm{HD}}}} = 280\,494$. For the outer and the inner iterations of the two-level iterative scheme, ${\chi_{{\mathrm{RRE}}}}= 10^{-4}$ and ${\chi_{{\mathrm{RRE}}}}= 10^{-8}$, respectively.

Fig.~\ref{fig_dimer}(b) plots the scattering cross section (SCS) computed from the solutions obtained using the two-level iterative scheme versus $\omega /{\omega _{\mathrm{p}}}$\cite{cao2016equivalent}. The transverse field resonance peaks are observed at $\omega = 0.6\,{\omega _{\mathrm{p}}} $ and $\omega = 0.75\,{\omega _{\mathrm{p}}}$ while the longitudinal field resonance peaks are observed at $\omega = 1.13\,{\omega_{\mathrm{p}}}$, $\omega = 1.29\,{\omega _{\mathrm{p}}}$, and $\omega = 1.50\,{\omega _{\mathrm{p}}}$. Furthermore, the figure shows that the nanodimer supports bonding and antibonding modes (generated by the transverse field resonances) at $\omega = 0.72\,{\omega _{\mathrm{p}}}$ and $\omega = 0.75\,{\omega _{\mathrm{p}}}$, respectively~\cite{zheng2018boundary}.
\begin{figure}[t!]
\centering
\subfigure[]{\includegraphics[width=0.6\columnwidth]{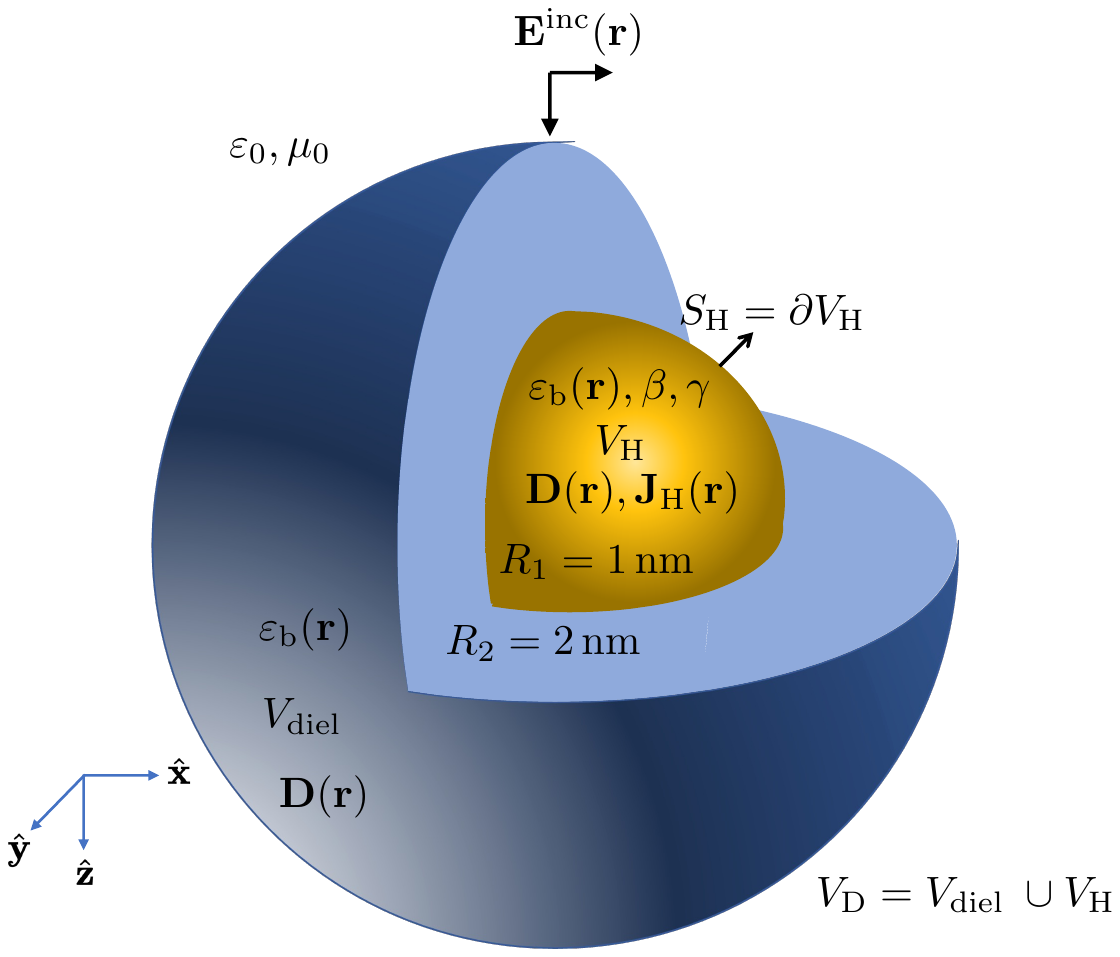}}\\
\subfigure[]{\includegraphics[width=0.6\columnwidth]{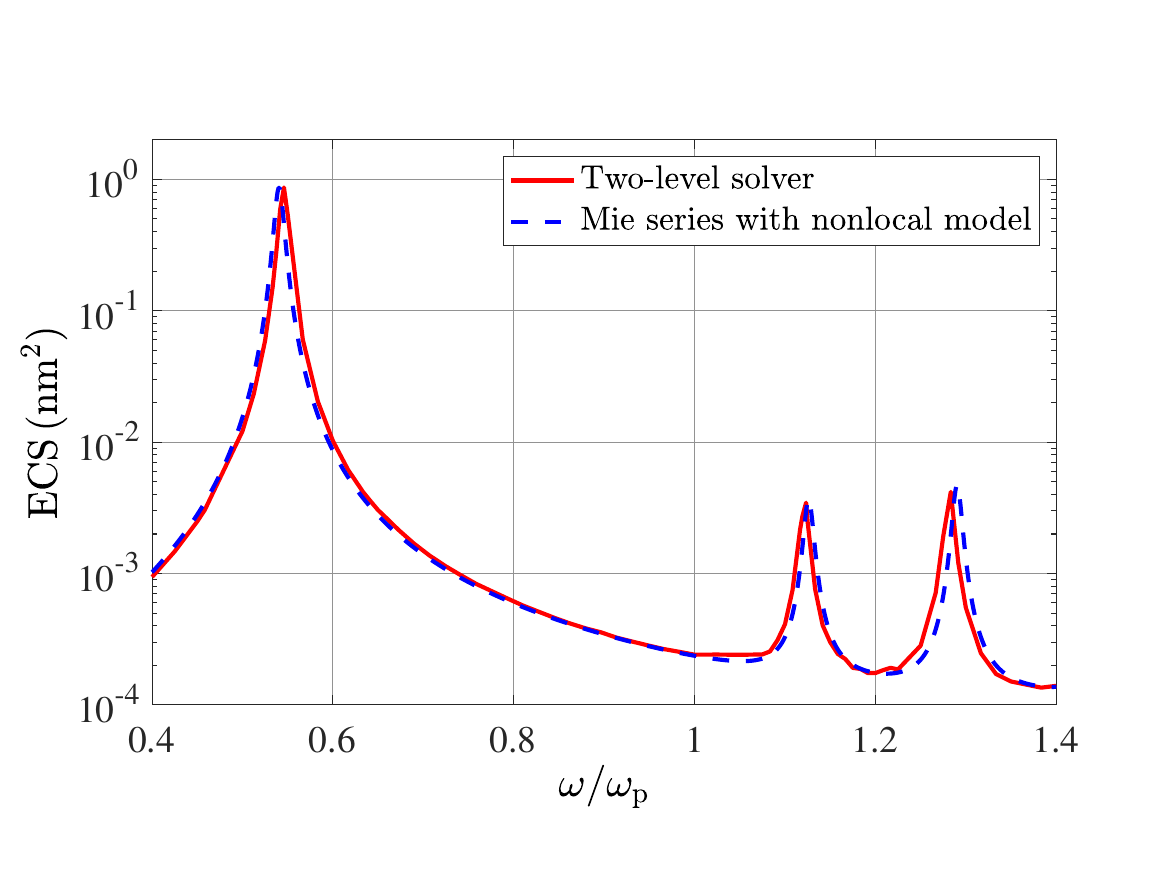}}\\
\caption{(a) Description of the scattering problem involving a silica-coated gold nanosphere. (b) ECS computed from the solutions obtained using the two-level iterative scheme and the Mie series with nonlocal-reponse material for the sphere and local-response material for the coating versus $\omega /{\omega _\mathrm{p}}$.}
\label{fig_composite}
\end{figure}
\subsection{Composite Sphere}
In this example, electromagnetic scattering from a silica-coated gold nanosphere [Fig.~\ref{fig_composite}(a)] is analyzed using the proposed solver. The radius of the gold sphere is $1\,\mathrm{nm}$ and the thickness of the silica coating is $1\,\mathrm{nm}$. The hydrodynamic equation parameters for gold are same as those in Section~\ref{sec:sphere} and the relative permittivity of silica is ${\varepsilon_{\mathrm{b}}(\mathbf{r})} = 2.25$. Note that for this problem, $V_{\mathrm{D}}=V_{\mathrm{diel}}\cup V_{\mathrm{H}}$.

A total of $60$ simulations are carried out using the two-level iterative scheme at equally spaced points in the frequency range $\omega \in [0.4 {\omega _{\mathrm{p}}}, 1.4 {\omega _{\mathrm{p}}}]$. In these simulations, $\mathbf{D}(\mathbf{r})$ induced inside the coating and the sphere and $ \mathbf{J}_{\mathrm{H}}(\mathbf{r})$ induced inside the sphere are discretized using ${N_{\mathrm{D}}} = 97\,642$ and ${N_{{\mathrm{H}}}} = 80\,486$ basis functions, respectively. For the outer and the inner iterations of the two-level iterative scheme, ${\chi_{{\mathrm{RRE}}}}= 10^{-4}$ and ${\chi_{{\mathrm{RRE}}}}= 10^{-8}$, respectively.

Fig.~\ref{fig_composite}(b) plots ECS computed from the solutions obtained by the two-level iterative scheme and the Mie series with nonlocal-response material model for the sphere and local-response material model for the coating versus $\omega /{\omega _{\mathrm{p}}}$. The results agree well, however, as expected, the error in the result obtained by the proposed method increases with $\omega$. This error can be reduced by using a denser mesh that can capture the behavior of the fields with large $k_{\mathrm{L}}(\mathbf{r})$ more accurately (see Section~\ref{sec:comm} and Fig.~\ref{fig:wavenum}).

The transverse field resonance peak is observed at  $\omega = 0.55\,{\omega _{\mathrm{p}}}$ and the longitudinal field resonance peaks are observed at $\omega = 1.13\,{\omega_{\mathrm{p}}}$ and $\omega = 1.31\,{\omega_{\mathrm{p}}}$. Comparing Figs.~\ref{fig_sphere}(c) and~\ref{fig_composite}(b), one can see that the transverse field resonance peak shifts towards lower frequencies due to the presence of the silica coating.

\begin{figure}[t!]
\centering
\subfigure[]{\includegraphics[width=0.6\columnwidth]{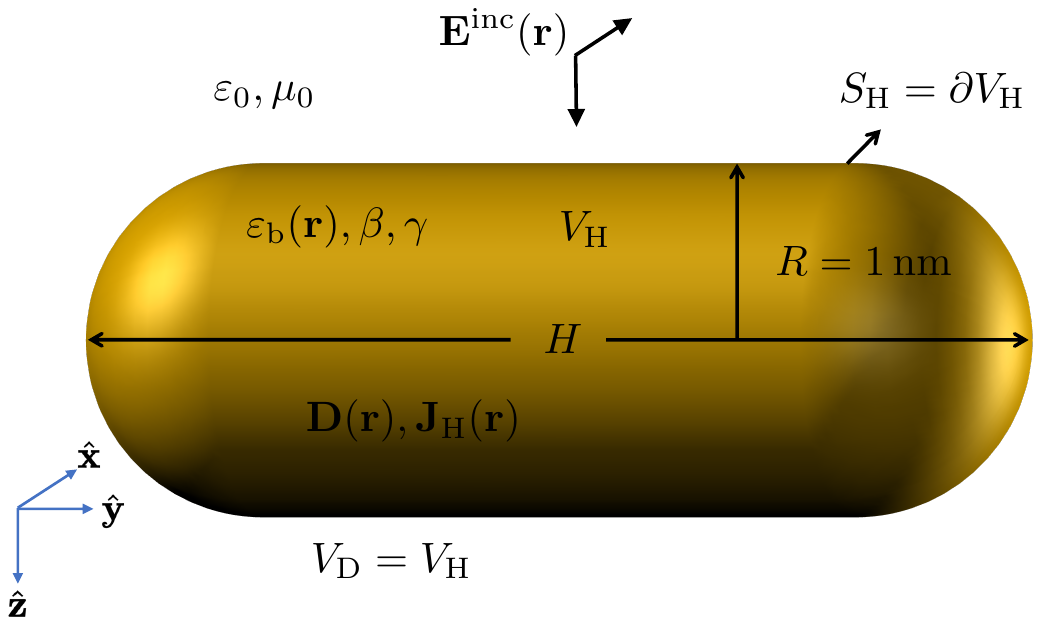}}\\
\subfigure[]{\includegraphics[width=0.6\columnwidth]{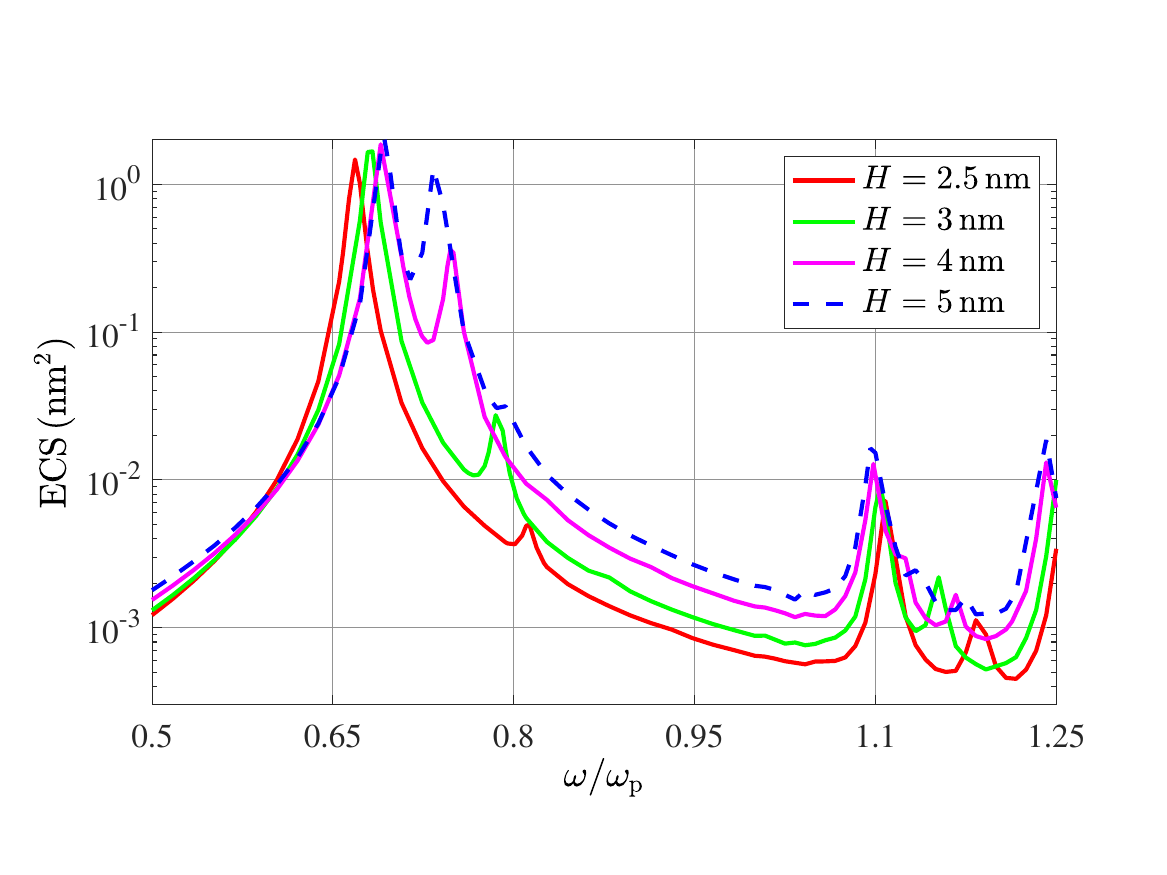}}\\
\caption{(a) Description of the scattering problem involving a gold nanocylinder. (b) ECS computed from the solutions obtained using the two-level iterative scheme for all the four nanocylinders versus $\omega /{\omega _\mathrm{p}}$.}
\label{fig_cylin}
\end{figure}

\begin{figure}[t!]
\centering
\subfigure[]{\includegraphics[width=0.6\columnwidth]{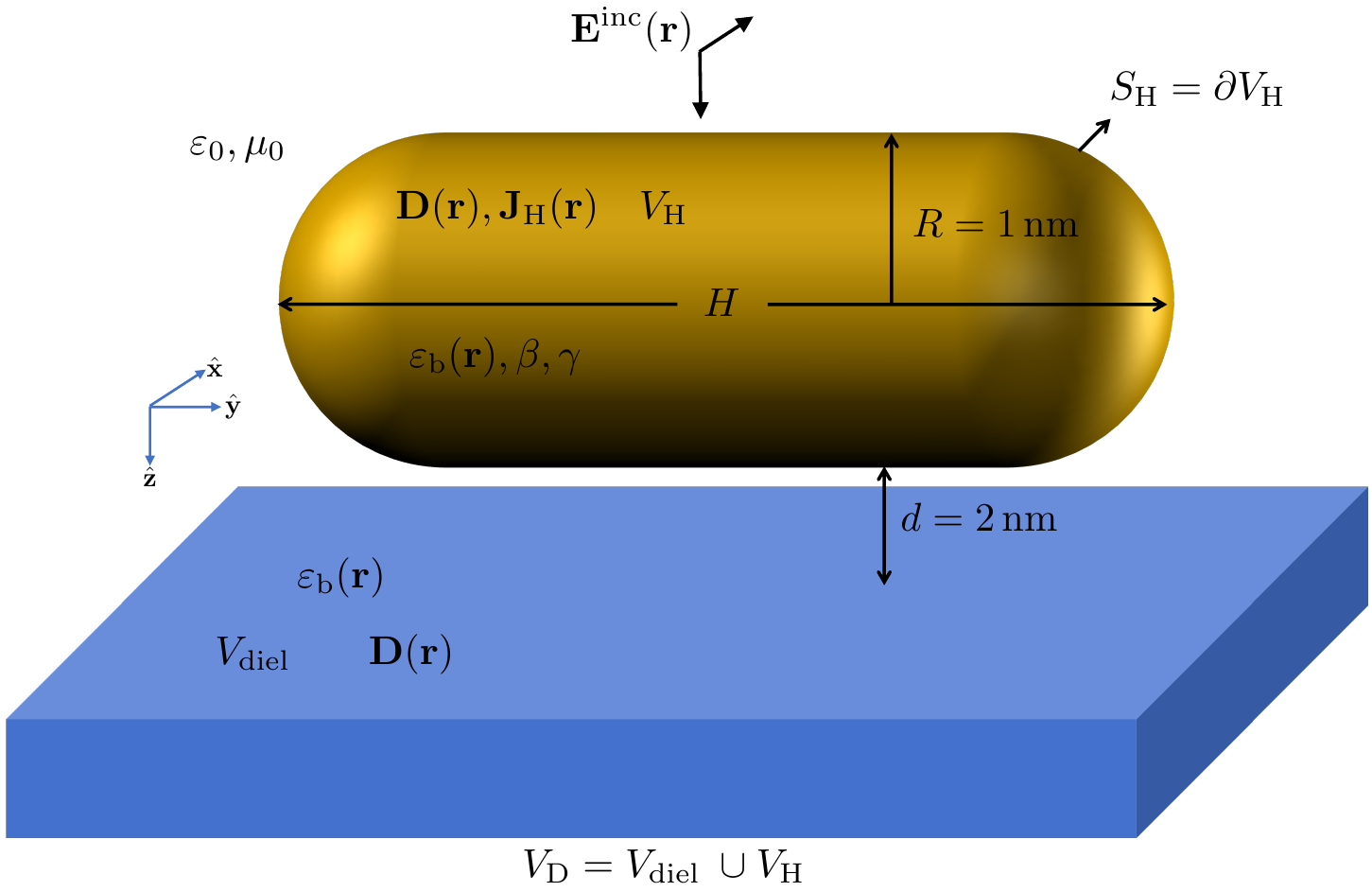}}\\
\subfigure[]{\includegraphics[width=0.6\columnwidth]{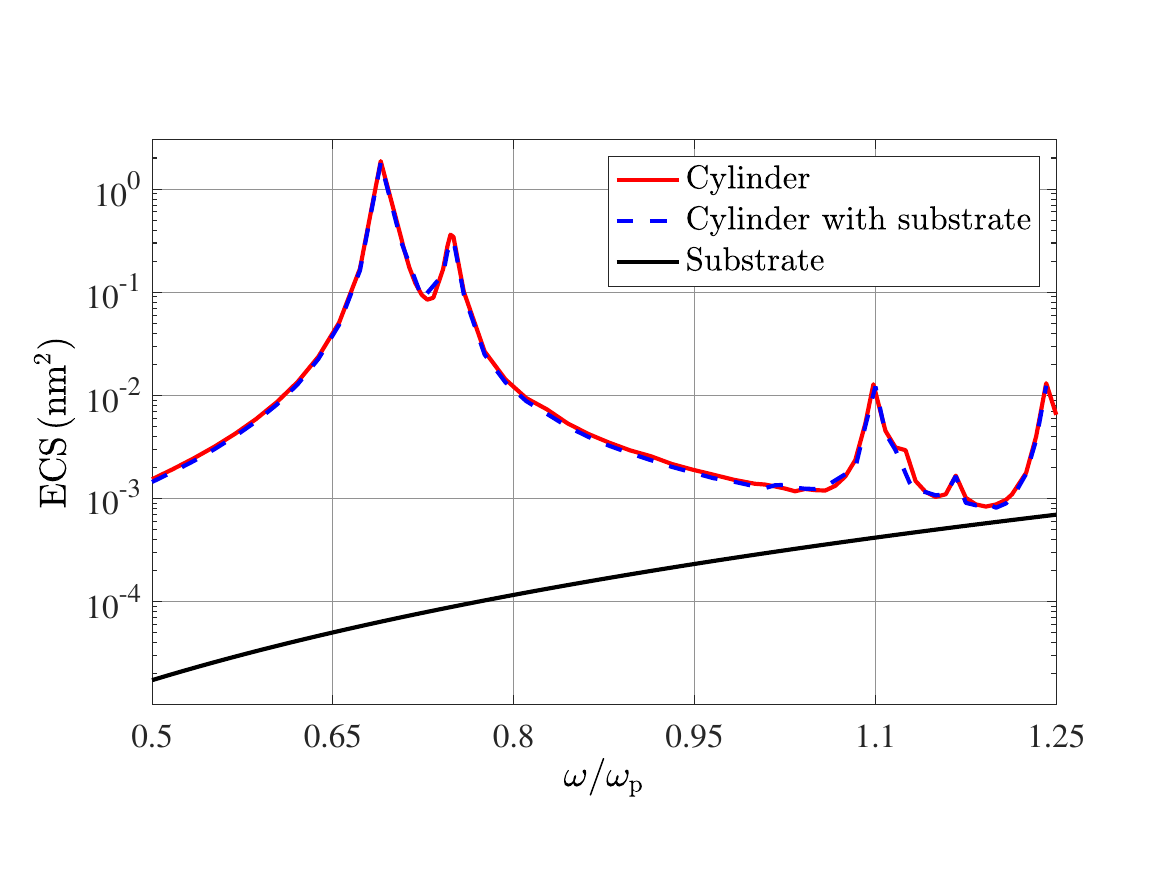}}\\
\caption{(a) Description of the scattering problem involving a gold nanocylinder on top of a silica cylinder. (b) ECS computed from the solutions obtained in all three scenarios (the scatterer consists of only the cylinder, the cylinder and the substrate, and only the substrate) versus $\omega /{\omega _\mathrm{p}}$.}
\label{fig_cylinslab}
\end{figure}
\subsection{Metallic Cylinder}\label{sec:cylinder}
In this example, electromagnetic scattering from a gold nanocylinder [Fig.~\ref{fig_cylin}(a)] is analyzed using the proposed method. Note that the ends of the cylinder are rounded here since sharp edges can not be often fabricated at nano-scales. Four nanocylinders with length ${H} \in \{2.5\,\mathrm{nm}, 3\,\mathrm{nm}, 4\,\mathrm{nm}, 5\,\mathrm{nm}\}$ are considered. All four cylinders have a radius of $1\,\mathrm{nm}$. The hydrodynamic equation parameters for gold are same as those in Section~\ref{sec:sphere}. Note that, for this problem, $ V_{\mathrm{D}}=V_{\mathrm{H}}$ and $ V_{\mathrm{diel}} = \emptyset$. 

A total of $60$ simulations are carried out using the two-level iterative scheme at equally spaced points in the frequency range $\omega \in [0.5 \omega_{\mathrm{p}}, 1.25 \omega_{\mathrm{p}}]$. Four levels of mesh are used for the four nanocylinders and these meshes use $ N_{\mathrm{D}} = \{58\,312, 60\,948, 61\,010, 62\,186\}$ and $ N_{\mathrm{H}} = \{56\,384, 58\,900, 58\,962, 60\, 006\}$ basis functions to discretize $\mathbf{D}(\mathbf{r})$ and $\mathbf{J}_{\mathrm{H}} (\mathbf{r}) $ induced inside the cylinder, respectively. For the outer and the inner iterations of the two-level iterative scheme, ${\chi_{{\mathrm{RRE}}}}= 10^{-4}$ and ${\chi_{{\mathrm{RRE}}}}= 10^{-8}$, respectively.

Fig.~\ref{fig_cylin}(b) compares ECS computed from the solutions obtained using the two-level iterative scheme for all the four nanocylinders versus $\omega /{\omega _{\mathrm{p}}}$. As seen from the figure, the transverse field resonance peak shifts from $\omega = 0.67\,{\omega _{\mathrm{p}}} $ to $\omega = 0.69\,{\omega _{\mathrm{p}}}$ when the length of the cylinder is increased from $2.5\,\mathrm{nm}$ to $5\,\mathrm{nm}$. The longitudinal field resonance peaks are observed at $\omega = 1.1\,{\omega_{\mathrm{p}}}$ and $\omega = 1.25\,{\omega _{\mathrm{p}}}$. In addition, this figure shows an extra resonance peak that shifts from $\omega = 0.81\,{\omega _{\mathrm{p}}}$ to $\omega = 0.73\,{\omega _{\mathrm{p}}}$ with increasing length. Note that this resonance (associated with the length of the cylinder along the $\hat{y}$-direction) is induced even though the incident electric field $\mathbf{E}^{\mathrm{inc}}(\mathbf{r})$ is polarized in the $\hat{x}$-direction. 

\subsection{Metallic Cylinder on top of a Dielectric Slab}\label{sec:cylin_slab}

For the last example, electromagnetic scattering from a gold nanocylinder located on top of a silica substrate [Fig.~\ref{fig_cylinslab}(a)] is analyzed using the proposed method. The length and the radius of the cylinder are $4\,\mathrm{nm}$ and $1\,\mathrm{nm}$, respectively. The width, the length, and the height of the slab are $ 6\,\mathrm{nm}$, $6\,\mathrm{nm}$, and $1\,\mathrm{nm}$, respectively. The shortest distance between the cylinder and the slab is $2\,\mathrm{nm}$. The hydrodynamic equation parameters for gold are same as those in Section~\ref{sec:sphere} and the relative electric permittivity of silica is $\varepsilon_{\mathrm{b}}(\mathbf{r})=2.25$. 

Three scattering scenarios are considered: (i) the scatterer is only the nanocylinder ($ V_{\mathrm{D}}=V_{\mathrm{H}}$ and $ V_{\mathrm{diel}} = \emptyset$, ${N_{\mathrm{D}}} = 61\,010$, and ${N_{{\mathrm{H}}}} = 58\,962$), (ii) the scatterer consists of the nanocylinder and the substrate ($V_{\mathrm{D}}=V_{\mathrm{diel}}\cup V_{\mathrm{H}}$, ${N_{\mathrm{D}}} = 61\,294$, and ${N_{{\mathrm{H}}}} = 58\,962$), and (iii) the scatterer is only the substrate ($V_{\mathrm{D}}=V_{\mathrm{diel}}$, $V_{\mathrm{H}} =\emptyset$, ${N_{\mathrm{D}}} = 284$, and ${N_{{\mathrm{H}}}} = 0$). Simulations of these three scenarios are carried out using the two-level iterative scheme [reduces to a ``traditional'' volume integral equation solver~\cite{botha2006,pasi2012} with single-level iterations for scenario (iii)] at $60$ equally spaced points in the frequency range $\omega \in [0.5 \omega_{\mathrm{p}}, 1.25 \omega_{\mathrm{p}}]$. Note that for scenario (ii) and (iii), only $284$ basis functions are used to discretize $\mathbf{D}({\mathbf{r}})$ induced inside the slab (see the discussion in Section~\ref{sec:comm}). For the outer and the inner iterations of the two-level iterative scheme, ${\chi_{{\mathrm{RRE}}}}= 10^{-4}$ and ${\chi_{{\mathrm{RRE}}}}= 10^{-8}$, respectively.

Fig.~\ref{fig_cylinslab}(b) plots ECS computed in the three scenarios described above versus $\omega /{\omega _{\mathrm{p}}}$. As expected, ECS of the silica slab [scenario (iii)] does not involve any resonance peaks since silica does not have any plasmonic properties. Secondly, the slab is located far away from the gold nanocylinder, so the coupling between them is not expected to be strong and ECS in scenario (i) and ECS in scenario (ii) are close to each other especially at the lower end of the frequency range. The difference increases at the higher end, which might be explained by the fact that ECS of the silica slab is larger at higher frequencies [scenario (iii)] and therefore contributes more to the total ECS in scenario (ii). 

\section{Conclusion}\label{sec:conc}
Electromagnetic scattering from nanostructures consisting of metallic and dielectric parts is analyzed by solving a coupled system of the volume integral and hydrodynamic equations. The hydrodynamic equation, which is enforced only in the metallic part, relates the free electron polarization current to the electric flux. This equation effectively updates the constitutive relation and permits modeling of the nonlocality. The volume integral equation, which is enforced in both the metallic and the dielectric parts, relates the electric flux to the scattered electric field. Unknown electric flux and free electron polarization current are expanded using a combination of full and half SWG basis functions. The boundary condition associated with the free electron polarization current on the metal-dielectric interface is enforced by excluding the half SWG basis functions, which are located on this interface, from the expansion of the free electron polarization current. Inserting these expansions into the coupled system of the volume integral and hydrodynamic equations and using Galerkin testing yield a matrix system.  

An efficient two-level iterative solver is developed to solve this matrix system. This approach inverts the discretized hydrodynamic equation (bottom rows of the matrix system) for the coefficients of the free electron polarization current and substitutes the result in the discretized volume integral equation (top rows of the matrix system). Outer iterations solve this reduced matrix system while the inner iterations invert the discretized hydrodynamic equation at every iteration of the outer iterations. 

Numerical experiments, which involve the computation of RCS, ECS, and SCS for metallic and composite nanostructures, are carried out to demonstrate the accuracy, the efficiency, and the applicability of the proposed method.

Future research directions include incorporation of different nonlocal hydrodynamic equations, implementation of boundary conditions on metal-metal interfaces, acceleration of the matrix solution using matrix compression schemes, and application of the proposed solver to different problems with different types of excitations.

% if have a single appendix:
\begin{appendices}
\section{Entries of the matrix and the right hand side vector in\texorpdfstring{~\eqref{eq14}}{ 12}}\label{sec:app}

While computing the entries of $\mathbf{Z}_{\mathrm{DD}}$, $\mathbf{Z}_{\mathrm{DH}}$, $\mathbf{Z}_{\mathrm{HD}}$, and $\mathbf{Z}_{\mathrm{HH}}$, it is assumed that $\varepsilon_b({\mathbf{r}})$ and $\kappa({\mathbf{r}})$ are constant in a given tetrahedron and the values of these constants are obtained by sampling $\varepsilon_b({\mathbf{r}})$ and $\kappa({\mathbf{r}})$ at the center of that tetrahedron. Therefore, one can define local functions as 
\begin{equation*}
\varepsilon_{\mathrm{b},n}(\mathbf{r}) =\left\{\begin{aligned}
&\varepsilon_{\mathrm{b},n}^{+} = \varepsilon_{\mathrm{b}}(\mathbf{r}_{\mathrm{c}}^{+}), \; \mathbf{r} \in V_{n}^{+} \\
&\varepsilon_{\mathrm{b},n}^{-} = \varepsilon_{\mathrm{b}}(\mathbf{r}_{\mathrm{c}}^{-}), \; \mathbf{r} \in V_{n}^{-}
\end{aligned}\right.
\end{equation*}
\begin{equation*}
\kappa_n(\mathbf{r}) =\left\{\begin{aligned}
&\kappa_n^{+} = \kappa(\mathbf{r}_{\mathrm{c}}^{+}), \; \mathbf{r} \in V_{n}^{+} \\
&\kappa_n^{-} = \kappa(\mathbf{r}_{\mathrm{c}}^{-}), \; \mathbf{r} \in V_{n}^{-}
\end{aligned}\right.
\end{equation*}
where $\mathbf{r}_{\mathrm{c}}^{\pm}$ are the centers of $V_n^{\pm}$.

%======================================
\subsection{Entries of \texorpdfstring{$\mathbf{Z}_{\mathrm{DD}}$}{ZDD}}
If both ${\mathbf{f}}_m^{\mathrm{D}}(\mathbf{r})$ and ${\mathbf{f}}_n^{\mathrm{D}}(\mathbf{r})$ are full SWG functions, then
\begin{equation*}
\begin{aligned}
&\{\mathbf{Z}_{\mathrm{DD}}\}_{m n}=\frac{1}{\varepsilon_{0}} \int_{V_{m}} \frac{\mathbf{f}_{m}^{\mathrm{D}}(\mathbf{r}) \cdot \mathbf{f}_{n}^{\mathrm{D}}(\mathbf{r})}{\varepsilon_{\mathrm{b},n}(\mathbf{r})} dv\\
&-\omega^{2} \mu_{0} \int_{V_{m}} \mathbf{f}_{m}^{\mathrm{D}}(\mathbf{r}) \cdot \int_{V_{n}} \kappa_{n}(\mathbf{r}^{\prime}) \mathbf{f}_{n}^{\mathrm{D}}(\mathbf{r}^{\prime}) G(\mathbf{r}, \mathbf{r}^{\prime}) dv^{\prime} dv \\
&+\frac{1}{\varepsilon_{0}} \int_{V_{m}} \nabla \cdot \mathbf{f}_{m}^{\mathrm{D}}(\mathbf{r})\left\{\int_{V_{n}} \kappa_n(\mathbf{r}^{\prime}) \nabla^{\prime} \cdot \mathbf{f}_{n}^{\mathrm{D}}(\mathbf{r}^{\prime}) G(\mathbf{r}, \mathbf{r}^{\prime}) dv^{\prime}\right. \\
&-(\kappa^{+}_n-\kappa^{-}_n)\left.\int_{S_{n}}G(\mathbf{r}, \mathbf{r}^{\prime}) ds^{\prime}\right\}dv.
\end{aligned}
\end{equation*}
If ${\mathbf{f}}_m^{\mathrm{D}}(\mathbf{r})$ is a half SWG function and ${\mathbf{f}}_n^{\mathrm{D}}(\mathbf{r})$ is a full SWG function, then 
\begin{equation*}
\begin{aligned}
&\{\mathbf{Z}_{\mathrm{DD}}\}_{m n}=\frac{1}{\varepsilon_{0}} \int_{V_{m}^{+}} \frac{\mathbf{f}_{m}^{\mathrm{D}}(\mathbf{r}) \cdot \mathbf{f}_{n}^{\mathrm{D}}(\mathbf{r})}{\varepsilon_{\mathrm{b},n}(\mathbf{r})}dv\\
&-\omega^{2} \mu_{0} \int_{V_{m}^{+}} \mathbf{f}_{m}^{\mathrm{D}}(\mathbf{r}) \cdot \int_{V_{n}} \kappa_n(\mathbf{r}^{\prime}) \mathbf{f}_{n}^{\mathrm{D}}(\mathbf{r}^{\prime}) G(\mathbf{r}, \mathbf{r}^{\prime}) dv^{\prime} dv \\
&+\frac{1}{\varepsilon_{0}} \int_{V_{m}^{+}} \nabla \cdot \mathbf{f}_{m}^{\mathrm{D}}(\mathbf{r})\left\{\int_{V_{n}} \kappa_n(\mathbf{r}^{\prime}) \nabla^{\prime} \cdot \mathbf{f}_{n}^{\mathrm{D}}(\mathbf{r}^{\prime}) G(\mathbf{r}, \mathbf{r}^{\prime}) dv^{\prime}\right.\\
&\left.-(\kappa_n^{+}-\kappa_n^{-})\int_{S_{n}} G(\mathbf{r}, \mathbf{r}^{\prime}) ds^{\prime}\right\}dv\\
&-\frac{1}{\varepsilon_{0}} \int_{S_{m}} \hat{\mathbf{n}}_m(\mathbf{r}) \cdot \mathbf{f}_{m}^{\mathrm{D}}(\mathbf{r})\left\{\int_{V_{n}} \kappa_n(\mathbf{r}^{\prime}) \nabla^{\prime} \cdot \mathbf{f}_{n}^{\mathrm{D}}(\mathbf{r}^{\prime}) G(\mathbf{r}, \mathbf{r}^{\prime})dv^{\prime}\right.\\
&\left.-(\kappa_n^{+}-\kappa_n^{-})\int_{S_{n}}G(\mathbf{r}, \mathbf{r}^{\prime}) ds^{\prime}\right\} ds
\end{aligned}
\end{equation*}
Here,  $\hat{\mathbf{n}}_m(\mathbf{r})$ is the unit normal vector on $S_m$ pointing from $V_n{-}$ to $V_n{+}$ and $\hat{\mathbf{n}}_{m}(\mathbf{r}) \cdot \mathbf{f}_{m}^{\mathrm{D}}(\mathbf{r})=1$.
If ${\mathbf{f}}_m^{\mathrm{D}}(\mathbf{r})$ is a full SWG function and ${\mathbf{f}}_n^{\mathrm{D}}(\mathbf{r})$ is a half SWG function, then
\begin{equation*}
\begin{aligned}
&\{\mathbf{Z}_{\mathrm{DD}}\}_{m n}=\frac{1}{\varepsilon_{0}{\varepsilon_{\mathrm{b},n}^{+}}}\int_{V_{m}}{\mathbf{f}_{m}^{\mathrm{D}}(\mathbf{r}) \cdot \mathbf{f}_{n}^{\mathrm{D}}(\mathbf{r})}dv\\
&-\omega^{2} \mu_{0}\kappa_{n}^{+}\int_{V_{m}} \mathbf{f}_{m}^{\mathrm{D}}(\mathbf{r}) \cdot \int_{V_{n}^{+}}\mathbf{f}_{n}^{\mathrm{D}}(\mathbf{r}^{\prime}) G(\mathbf{r}, \mathbf{r}^{\prime}) dv^{\prime} dv \\
&+\frac{\kappa_n^{+}}{\varepsilon_{0}} \int_{V_{m}} \nabla \cdot \mathbf{f}_{m}^{\mathrm{D}}(\mathbf{r})\left\{\int_{V_{n}^{+}}\nabla^{\prime} \cdot \mathbf{f}_{n}^{\mathrm{D}}(\mathbf{r}^{\prime}) G(\mathbf{r}, \mathbf{r}^{\prime}) dv^{\prime}\right. \\
&-\left.\int_{S_{n}}G(\mathbf{r}, \mathbf{r}^{\prime}) ds^{\prime}\right\}dv.
\end{aligned}
\end{equation*}
If both ${\mathbf{f}}_m^{\mathrm{D}}(\mathbf{r})$ and ${\mathbf{f}}_n^{\mathrm{D}}(\mathbf{r})$ are half SWG functions, then
\begin{equation*}
\begin{aligned}
&\{\mathbf{Z}_{\mathrm{DD}}\}_{m n}=\frac{1}{{\varepsilon_{0}\varepsilon_{\mathrm{b},n}^{+}}}\int_{V_{m}^{+}}{\mathbf{f}_{m}^{\mathrm{D}}(\mathbf{r}) \cdot \mathbf{f}_{n}^{\mathrm{D}}(\mathbf{r})}dv\\
&-\omega^{2} \mu_{0}\kappa_n^{+}\int_{V_{m}^{+}} \mathbf{f}_{m}^{\mathrm{D}}(\mathbf{r}) \cdot \int_{V_{n}^{+}} \mathbf{f}_{n}^{\mathrm{D}}(\mathbf{r}^{\prime}) G(\mathbf{r}, \mathbf{r}^{\prime}) dv^{\prime} dv \\
&+\frac{\kappa_n^{+}}{\varepsilon_{0}} \int_{V_{m}^{+}} \nabla \cdot \mathbf{f}_{m}^{\mathrm{D}}(\mathbf{r})\left\{\int_{V_{n}^{+}}\nabla^{\prime} \cdot \mathbf{f}_{n}^{\mathrm{D}}(\mathbf{r}^{\prime}) G(\mathbf{r}, \mathbf{r}^{\prime}) dv^{\prime}\right.\\
&\left.-\int_{S_{n}} G(\mathbf{r}, \mathbf{r}^{\prime}) ds^{\prime}\right\}dv\\
&-\frac{\kappa_n^{+}}{\varepsilon_{0}} \int_{S_{m}} \hat{\mathbf{n}}_m(\mathbf{r}) \cdot \mathbf{f}_{m}^{\mathrm{D}}(\mathbf{r})\left\{\int_{V_{n}^{+}}\nabla^{\prime} \cdot \mathbf{f}_{n}^{\mathrm{D}}(\mathbf{r}^{\prime}) G(\mathbf{r}, \mathbf{r}^{\prime})dv^{\prime}\right.\\
&\left.-\int_{S_{n}}G(\mathbf{r}, \mathbf{r}^{\prime}) ds^{\prime}\right\} ds.
\end{aligned}
\end{equation*}

%======================================
\subsection{Entries of \texorpdfstring{$\mathbf{Z}_{\mathrm{DH}}$}{ZDH}}
${\mathbf{f}}_n^{\mathrm{H}}(\mathbf{r})$ is always a full SWG function. If ${\mathbf{f}}_m^{\mathrm{D}}(\mathbf{r})$ is a full SWG function, then
\begin{equation*}
\begin{aligned}
&\{\mathbf{Z}_{\mathrm{DH}}\}_{m n}=-\frac{1}{j \omega\varepsilon_{0}} \int_{V_{m}} \frac{\mathbf{f}_{m}^{\mathrm{D}}(\mathbf{r}) \cdot \mathbf{f}_{n}^{\mathrm{H}}(\mathbf{r})}{\varepsilon_{\mathrm{b},n}(\mathbf{r})} dv\\
&+j \omega \mu_{0} \int_{V_{m}} \mathbf{f}_{m}^{\mathrm{D}}(\mathbf{r}) \cdot \int_{V_{n}} \frac{\mathbf{f}_{n}^{\mathrm{H}}(\mathbf{r})}{\varepsilon_{\mathrm{b},n}(\mathbf{r}^{\prime})} G(\mathbf{r}, \mathbf{r}^{\prime}) dv^{\prime} dv \\
&+\frac{1}{j \omega \varepsilon_{0}} \int_{V_{m}} \nabla \cdot \mathbf{f}_{m}^{\mathrm{D}}(\mathbf{r})\left\{\int_{V_{n}} \frac{\nabla^{\prime} \cdot \mathbf{f}_{n}^{\mathrm{H}}(\mathbf{r}^{\prime}) }{\varepsilon_{\mathrm{b},n}(\mathbf{r}^{\prime})}G(\mathbf{r},\mathbf{r}^{\prime}) dv^{\prime}\right.\\
&\left.-\left[\frac{1}{\varepsilon_{\mathrm{b},n}^{+}}-\frac{1}{\varepsilon_{\mathrm{b},n}^{-}}\right]\int_{S_{n}}G(\mathbf{r}, \mathbf{r}^{\prime}) ds^{\prime}\right\}dv.
\end{aligned}
\end{equation*}
If ${\mathbf{f}}_m^{\mathrm{D}}(\mathbf{r})$ is a half SWG function, then 
\begin{equation*}
\begin{aligned}
&\{\mathbf{Z}_{\mathrm{DH}}\}_{m n}=-\frac{1}{j \omega\varepsilon_{0}} \int_{V_{m}^{+}} \frac{\mathbf{f}_{m}^{\mathrm{D}}(\mathbf{r}) \cdot \mathbf{f}_{n}^{\mathrm{H}}(\mathbf{r})}{\varepsilon_{\mathrm{b},n}(\mathbf{r})}dv \\
&+j \omega \mu_{0} \int_{V_{m}^{+}} \mathbf{f}_{m}^{\mathrm{D}}(\mathbf{r}) \cdot \int_{V_{n}} \frac{\mathbf{f}_{n}^{\mathrm{H}}(\mathbf{r})}{\varepsilon_{\mathrm{b},n}(\mathbf{r}^{\prime})} G(\mathbf{r}, \mathbf{r}^{\prime}) dv^{\prime} dv \\
&+\frac{1}{j \omega \varepsilon_{0}} \int_{V_{m}^{+}} \nabla \cdot \mathbf{f}_{m}^{\mathrm{D}}(\mathbf{r})\left\{\int_{V_{n}} \frac{\nabla^{\prime} \cdot \mathbf{f}_{n}^{\mathrm{H}}(\mathbf{r}^{\prime}) }{\varepsilon_{\mathrm{b},n}(\mathbf{r}^{\prime})}G(\mathbf{r},\mathbf{r}^{\prime}) dv^{\prime}\right.\\
&\left.-\left[\frac{1}{\varepsilon_{\mathrm{b},n}^{+}}-\frac{1}{\varepsilon_{\mathrm{b},n}^{-}}\right]\int_{S_{n}}G(\mathbf{r}, \mathbf{r}^{\prime}) ds^{\prime}\right\}dv\\
&-\frac{1}{j \omega \varepsilon_{0}} \int_{S_{m}} \hat{\mathbf{n}}_{m}(\mathbf{r}) \cdot \mathbf{f}_{m}^{\mathrm{D}}(\mathbf{r})\left\{\int_{V_{n}} \frac{\nabla^{\prime} \cdot \mathbf{f}_{n}^{\mathrm{H}}(\mathbf{r}^{\prime}) }{\varepsilon_{\mathrm{b},n}(\mathbf{r}^{\prime})}G(\mathbf{r},\mathbf{r}^{\prime}) dv^{\prime}\right.\\
&\left.-\left[\frac{1}{\varepsilon_{\mathrm{b},n}^{+}}-\frac{1}{\varepsilon_{\mathrm{b},n}^{-}}\right]\int_{S_{n}}G(\mathbf{r}, \mathbf{r}^{\prime}) ds^{\prime}\right\}ds.
\end{aligned}
\end{equation*}
%======================================
\subsection{Entries of \texorpdfstring{$\mathbf{Z}_{\mathrm{HD}}$}{ZHD}}
${\mathbf{f}}_m^{\mathrm{H}}(\mathbf{r})$ is always a full SWG function. If ${\mathbf{f}}_n^{\mathrm{D}}(\mathbf{r})$ is a full SWG function, then 
\begin{equation*}
 \{{{{\mathbf{Z}}_{\mathrm{HD}}}}\}_{mn} = j\omega\omega^2_{\mathrm{p}}\int_{V_m} \frac{{{{\mathbf{f}}_m^{\mathrm{H}}}({\mathbf{r}}) \cdot {{\mathbf{f}}_n^{\mathrm{D}}}({\mathbf{r}})}}{{{\varepsilon_{\mathrm{b},n}(\mathbf{r})}}}dv.
\end{equation*}
If ${\mathbf{f}}_n^{\mathrm{D}}(\mathbf{r})$ is a half SWG function, then 
\begin{equation*}
\{{{{\mathbf{Z}}_{\mathrm{HD}}}}\}_{mn} = \frac{j\omega \omega^2_{\mathrm{p}}}{{{\varepsilon^{+}_{\mathrm{b},n}}}}\int_{V_m} {{{\mathbf{f}}_m^{\mathrm{H}}}({\mathbf{r}}) \cdot {{\mathbf{f}}_n^{\mathrm{D}}}({\mathbf{r}})}dv.
\end{equation*}
%======================================
\subsection{Entries of \texorpdfstring{$\mathbf{Z}_{\mathrm{HH}}$}{ZHH}}
${\mathbf{f}}_m^{\mathrm{H}}(\mathbf{r})$ and ${\mathbf{f}}_n^{\mathrm{H}}(\mathbf{r})$ are always full SWG functions:
\begin{equation*}
\begin{aligned}
\{\mathbf{Z}_{\mathrm{HH}}\}_{m n} &=-\beta^{2} \int_{V_{m}} [\nabla \cdot \mathbf{f}_{m}^{\mathrm{H}}(\mathbf{r})] [\nabla \cdot \mathbf{f}_{n}^{\mathrm{H}}(\mathbf{r})] dv\\
&+\omega(\omega-j \gamma) \int_{V_{m}} \mathbf{f}_{m}^{\mathrm{H}}(\mathbf{r}) \cdot \mathbf{f}_{n}^{\mathrm{H}}(\mathbf{r}) dv \\
&- \omega_{\mathrm{p}}^{2} \int_{V_{m}} \frac{\mathbf{f}_{m}^{\mathrm{H}}(\mathbf{r}) \cdot \mathbf{f}_{n}^{\mathrm{H}}(\mathbf{r})}{\varepsilon_{\mathrm{b},n}(\mathbf{r})} dv.
\end{aligned}
\end{equation*}
\subsection{Entries of \texorpdfstring{$\mathbf{V}^{\mathrm{inc}}$}{Vinc}} 
If ${\mathbf{f}}_m^{\mathrm{D}}$ is a full SWG function, then
\begin{equation*}
\{{\mathbf{V}}^{\mathrm{inc}}\}_m = \int_{V_m} {{\mathbf{f}}_m^{\mathrm{D}}}({\mathbf{r}}) \cdot {{\mathbf{E}}^{{\mathrm{inc}}}}({\mathbf{r}})dv.  
\end{equation*}
If ${\mathbf{f}}_m^{\mathrm{D}}$ is a half SWG function, then
\begin{equation*}
\{{\mathbf{V}}^{\mathrm{inc}}\}_m = \int_{V_m^{+}} {{\mathbf{f}}_m^{\mathrm{D}}}({\mathbf{r}}) \cdot {{\mathbf{E}}^{{\mathrm{inc}}}}({\mathbf{r}})dv.  
\end{equation*}
\end{appendices}
\newpage

% Generated by IEEEtran.bst, version: 1.14 (2015/08/26)

\end{document}